\documentclass[aps,
prd,
twocolumn,
superscriptaddress,
nobibnotes,
nofootinbib,
longbibliography,
preprintnumbers,
floatfix,
10pt]{revtex4-2}

\usepackage{graphicx}
\usepackage[dvipsnames,svgnames]{xcolor}
\usepackage{amsmath,amsfonts,amsthm,amssymb}
\usepackage[colorlinks]{hyperref}
\usepackage{mathtools}
\usepackage{bm}
\usepackage{multirow}
\usepackage[normalem]{ulem}
\usepackage{float}
\usepackage{booktabs}
\usepackage{microtype}
\usepackage{titlesec}
\usepackage{soul}
\usepackage{isotope}
\usepackage{xspace}
\usepackage[noabbrev,capitalise]{cleveref}
\makeatletter
\AddToHook{cmd/appendix/before}{\def\cref@section@alias{appendix}}
\makeatother

\graphicspath{{./figures/}}

\newcommand{\orcid}[1]{\href{https://orcid.org/#1}{\includegraphics[width=10pt]{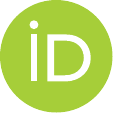}}}

\hypersetup{
    pdfnewwindow=true,      
    colorlinks=true,       
    linkcolor=violet,          
    citecolor=violet,        
    filecolor=violet,      
    urlcolor=violet        
}

\newcommand{\nuebar}{\bar{\nu}_e}
\newcommand{\thone}{\theta_{13}}
\newcommand{\thtwo}{\theta_{12}}
\newcommand{\thtwothree}{\theta_{23}}
\newcommand{\dmee}{\Delta m^2_{ee}}
\newcommand{\dmtwo}{\Delta m^2_{21}}
\newcommand{\Dee}{\Delta_{ee}}
\newcommand{\Dtwo}{\Delta_{21}}
\newcommand{\sinsqtwo}[1]{\sin^2(2#1)}
\newcommand{\relsth}{\sigma(\sinsqtwo{\thone})/\sinsqtwo{\thone}}
\newcommand{\eV}{\,\text{eV}}
\newcommand{\MeV}{\,\text{MeV}}
\newcommand{\km}{\,\text{km}}
\newcommand{\mwth}{\,\text{MW}_{\text{th}}}
\newcommand{\gwth}{\,\text{GW}_{\text{th}}}

\newcommand{\kilometers}{\,\text{km}}
\newcommand{\cubicmeters}{\,\text{m}^3}

\newcommand{\uthirtyeight}{${}^{238}$U\xspace}
\newcommand{\uthirtyfive}{${}^{235}$U\xspace}
\newcommand{\pthirtynine}{${}^{239}$Pu\xspace}
\newcommand{\pfortyone}{${}^{241}$Pu\xspace}

\newcommand{\ourtitle}{Future avenues for precision oscillation measurements with compact movable reactors} 

\begin{document}

\title{\ourtitle}

\author{Matheus Hostert \orcid{0000-0002-9584-8877}}
\email{matheus-hostert@uiowa.edu}
\affiliation{Department of Physics and Astronomy, \href{https://ror.org/036jqmy94}{University of Iowa}, Iowa City, IA 52242, USA}

\author{Stephan A. Meighen-Berger \orcid{0000-0001-6579-2000}\,}
\email{stephan-meighen-berger@uiowa.edu}
\affiliation{Department of Physics and Astronomy, \href{https://ror.org/036jqmy94}{University of Iowa}, Iowa City, IA 52242, USA}

\author{Maxim Pospelov}
\email{pospelov@umn.edu}
\affiliation{School of Physics and Astronomy, \href{https://ror.org/017zqws13}{University of Minnesota}, Minneapolis, MN 55455, USA}
\affiliation{William I. Fine Theoretical Physics Institute, School of Physics and Astronomy, \href{https://ror.org/017zqws13}{University of Minnesota}, Minneapolis, MN 55455, USA}
\affiliation{Theoretical Physics Department, \href{https://ror.org/01ggx4157}{CERN}, 1211 Geneva, Switzerland}


\begin{abstract}
No planned neutrino experiment is expected to surpass Daya Bay’s precision on $\thone$.
Doing so requires innovative strategies to reduce statistical and systematic uncertainties.
In this study, we consider a futuristic program with a movable reactor antineutrino source placed within short distances ($L\lesssim 3\,\km$) from the Jiangmen Underground Neutrino Observatory (JUNO) detector. 
Due to the large fiducial mass of JUNO, even a relatively small reactor can produce a statistically significant number of events over a few years to probe the $\lesssim 8.7\% \,\nuebar$ disappearance set by $\thone$.
We consider two possibilities: a $\mathcal{O}(20\mwth)$ microreactor in alcoves of the access tunnel to the underground laboratory, and a $\mathcal{O}(400 \mwth)$ unit on the surface.
In both cases the reactor is observed by a small flux monitor detector near the source, while JUNO measures the antineutrino flux at two baselines, suppressing correlated flux and detector systematics.
Using the measured Daya Bay per-isotope spectra and its covariance, we model a high-assay low-enriched uranium (HALEU) core for the microreactor and a low-enriched uranium (LEU) core for the surface unit, and discuss various potential sources of systematic uncertainties and their impact on the precision of $\thone$.
The underground microreactor could improve on JUNO's reach by a factor of three to four, to $\relsth \simeq 4\%$ over a decade, while the surface reactor would reach the present global-fit precision, $\relsth=2.5\%$, in about three years, provided the emission rate is monitored across reactor relocations to the sub-percent level.
\end{abstract}

\maketitle



\section{Introduction}
\label{sec:introduction}

Over two decades after neutrino oscillations were established~\cite{Super-Kamiokande:1998kpq,SNO:2002tuh,KamLAND:2002uet}, the field has entered an era of precision: the mass-squared differences are known to about $1\%$ and $\sin^2\theta_{ij}$ to a few percent~\cite{Esteban:2024eli,Capozzi:2025wyn,deSalas:2020pgw}.
The two remaining unknowns of the three-neutrino picture are the mass ordering (MO) and $\delta_{\rm CP}$.
The MO will be determined by JUNO~\cite{Cabrera:2020ksc,JUNO:2015zny,Forero:2021lax} in combination with accelerator~\cite{Nunokawa:2005nx,Parke:2024xre,NOvA:2026tir} and atmospheric data~\cite{Arguelles:2022hrt}, and independently by DUNE~\cite{DUNE:2020lwj}, while $\delta_{\rm CP}$ will be measured by DUNE and Hyper-Kamiokande~\cite{DUNE:2020lwj,Hyper-Kamiokande:2018ofw}.
With JUNO projected to measure $\sin^2\thtwo$ to better than $0.5\%$~\cite{JUNO:2022mxj,JUNO:2025gmd}, and DUNE and Hyper-Kamiokande $\sin^2\thtwothree$ to $0.5$–$3\%$, depending on its true value, $\thone$ may very well become the least precisely known mixing angle in the next decade.

\begin{figure}[t]
\centering
\includegraphics[width=0.47\textwidth]{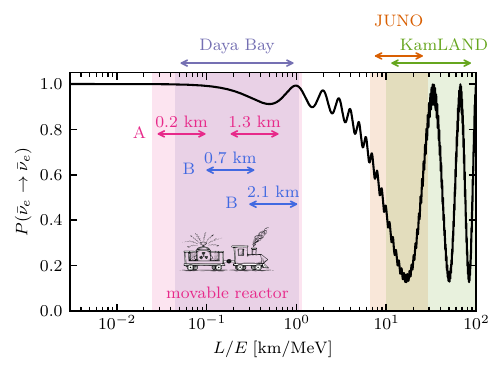}
\caption{
  Electron-antineutrino survival probability $P_{ee}$ as a function of $L/E$. 
  The first minimum reaches an $8.7\%$ depth set exclusively by $\thone$ near $L/E \simeq 0.5\,\km/\MeV$. 
  The solar mass splitting contribution to the deficit is of the per-mille level in the first minimum.
  A movable source would allow the fixed JUNO detector to sample several baselines in turn, scanning this curve in both baseline and energy.
  }
\label{fig:sketch}
\end{figure}

The current precision on $\thone$ is dominated by three dedicated multi-detector reactor experiments. 
Daya Bay achieved the world-leading $\sin^2\!2\thone$ constraint using eight approximately identical detectors at near ($\simeq 0.5\,\km$) and far ($\simeq 1.6\,\km$) sites around the Daya Bay and Ling Ao reactors~\cite{DayaBay:2007fgu,DayaBay:2018yms,DayaBay:2022orm}.
It established
\begin{align}\label{eq:DayaBay_Gd}
    \sinsqtwo{\thone} &= 0.0851  \times (1 \pm 2.8\%),
\\
     \frac{\dmee}{10^{-3}\eV^2 } &= 2.519 \times (1 \pm 2.4\%),
\end{align}
where $\Delta m^2_{ee} \equiv \cos^2\theta_{12}\,\Delta m_{31}^2 + \sin^2\theta_{12}\,\Delta m^2_{32}$~\cite{Nunokawa:2005nx,Parke:2016joa}.
The RENO experiment used a two-detector configuration at the Hanbit plant in South Korea, reaching a comparable but somewhat smaller sensitivity~\cite{RENO:2018dro}.
Double Chooz operated first with a far detector only and later added a near detector using the two Chooz reactors in les Ardennes in France~\cite{DoubleChooz:2019qbj}. 
Global fits currently quote the precision on $\sin^22\thone$ as $2.5\%$ ($1\sigma$)~\cite{Esteban:2024eli} and a recent analysis combining Daya Bay data with capture on gadolinium and hydrogen reaches~\cite{DayaBay:2024hrv},
\begin{equation}\label{eq:DayaBay_H}
    \sin^2(2\thone) = 0.0833 \times ( 1 \pm 2.6\%).
\end{equation} 
JUNO can measure $\thone$, but with a precision that is not competitive with Daya Bay because at $52.5\,\km$ the $\thone$-dependent signal is only a $\lesssim 2\%$ modulation of an already oscillated spectrum, far smaller than the $8.7\%$ dip at the first $\nuebar$ survival minimum at short baselines.
JUNO alone is projected to reach $\relsth \simeq 12\%$ after six years of data taking~\cite{JUNO:2022mxj}.
DUNE and Hyper-Kamiokande will not be competitive either.

\begin{figure*}[t]
\centering
\includegraphics[width=\textwidth]{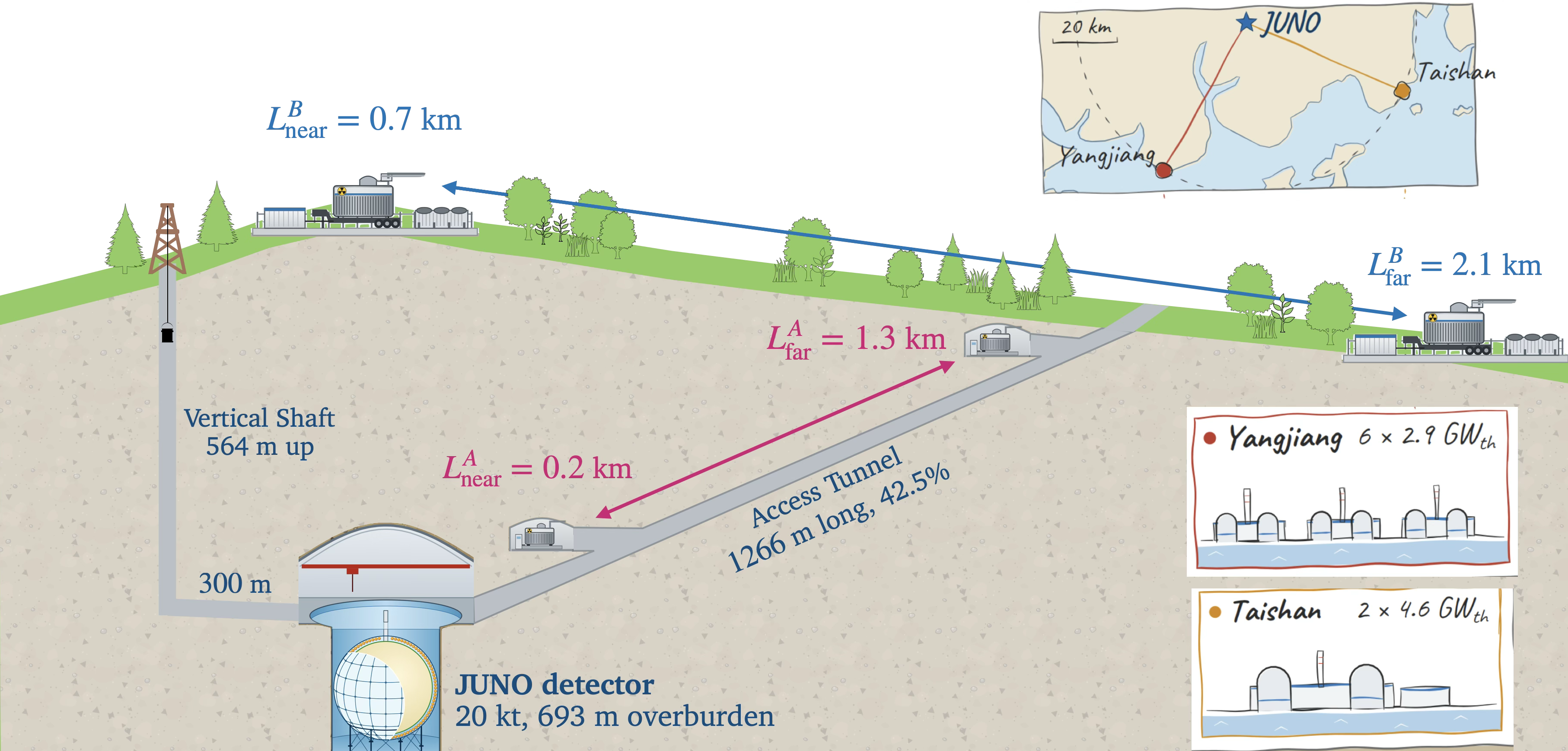}
\caption{A schematic of the JUNO detector and potential sites for movable compact reactors (not to scale).
\textcolor{purple}{\textbf{Configuration A}} microreactors are shown in alcoves along the access tunnel, and the \textcolor{RoyalBlue}{\textbf{Configuration B}} surface small modular reactors are placed on pads above ground.
Some of the existing far GW-scale reactor complexes are shown in the inset for reference, as these would continue to operate and serve as a background to the $\thone$ measurement.
}
\label{fig:sketch_2}
\end{figure*}

There are several benefits to reducing the uncertainty on $\thone$.
Since the leading term of the $\nu_\mu\to\nu_e$ appearance probability scales as $\sin^2\thtwothree\sin^2 2\thone$, an independent $\thone$ measurement improves the octant determination near maximal mixing ($\theta_{23}\simeq \pi/4$).
In addition, comparing reactor and accelerator determinations of $\thone$ tests the unitarity of the leptonic mixing matrix~\cite{Qian:2013ora,Parke:2015goa,Ellis:2020hus}, and this method may be limited by the reactor uncertainty within a decade.
The smallest factor of the Jarlskog invariant, $J_{\rm CP} = c_{12}s_{12}c_{23}s_{23}c^2_{13}s_{13}\sin\delta_{\rm CP} \simeq 0.033\sin\delta_{\rm CP}$~\cite{Jarlskog:1985ht}, is $s_{13}$, so $\sigma(J_{\rm CP})/J_{\rm CP} \supset \sigma(\thone)/\thone$, although a better reactor prior only mildly improves the $\delta_{\rm CP}$ reach of DUNE and Hyper-Kamiokande~\cite{Nunokawa:2007qh,Barger:2001yr,DUNE:2020lwj}.
Finally, $\thone$ has been a key discriminant of flavor-symmetry models, whose sum rules relate it to $\thtwo$, $\thtwothree$ and $\delta_{\rm CP}$~\cite{King:2013eh}.
Future oscillation measurements will significantly reduce the number of viable flavor models~\cite{Denton:2026wpy} and $\thone$ could help narrow down their underlying parameter space further.

Daya Bay was the result of a monumental experimental effort. 
It sampled antineutrinos from six reactor cores of 2.9~$\gwth$ each, using eight identically designed detectors in two near halls and one far hall. 
The near–far arrangement made reactor-flux and fuel-composition uncertainties largely cancel in relative comparisons. 
Each detector contained about 20 tonnes of liquid scintillator with 0.1\% gadolinium doping. 
The collaboration had already observed $\theta_{13}$-driven disappearance at more than $5\sigma$ with about 50 days of data in 2012~\cite{DayaBay:2012fng}, before the full 8-detector setup was even installed. 
It ran for nearly nine years, collecting 3158 days of data between December 2011 and December 2020 and a total of $5.55\times 10^6$ inverse beta decay (IBD) events~\cite{DayaBay:2022orm}.
That the precision in \cref{eq:DayaBay_Gd,eq:DayaBay_H} is not expected to be surpassed by any currently running or approved experiment illustrates both the power of the Daya Bay design and how difficult further improvement of $\theta_{13}$ will be with conventional reactor setups.
Only a few proposals to achieve that exist such as SuperChooz that would utilize the past underground cavern for the dismantled Chooz A reactor in les Ardennes~\cite{cabrera_serra_2022_7504162} and improved scintillation technology for cosmic background rejection explored by the LiquidO project~\cite{LiquidO:2019mxd}.
Other more ambitious reactor setups to achieve sub-percent precision have also been recently discussed~\cite{Zhang:2022zoc}.

In this study, we turn to the possibility of using JUNO for an ambitious post-mass-ordering program: monitoring \textit{movable} nuclear-reactor(s) to collect an unprecedented number of IBD and elastic scattering events within a single detector.
In this first study, we explore different compact reactor setups with baselines of $L\simeq 0.2$ to $3\,\km$ as shown in \cref{fig:sketch}.
These can exploit JUNO's large mass ($20\,$kton), excellent energy resolution (${\sim}3\%$ at $1\,\MeV$~\cite{JUNO:2024fdc}), and low-background environment to probe the $L/E$ dependence of atmospheric-frequency $\Delta m^2_{31}$ oscillations.
Due to their proximity, even relatively low-power reactors can overwhelm the signal of JUNO's fixed $\gwth$-scale far reactors.
By moving such reactors, one can invert the strategy of Daya Bay, RENO, and Double Chooz, and cancel correlated flux and spectral systematics by comparing near and far placements of the reactor with a single fixed detector.
Continuous operation of reactors would be preferable, but restarts of the core or replica reactors are more realistic on land and if closely monitored, may also achieve the desired effect.

For this study we focus on two potential configurations with different logistical constraints on reactor power as illustrated in \cref{fig:sketch_2}:
\begin{enumerate}
\item[]\textbf{Configuration~A}: a $\lesssim 20\mwth$ microreactor placed in alcoves of the access tunnel to the underground laboratory (or buried in the surrounding rock), located $0.2 - 1.3\,\km$ from the JUNO detector, offering a measurement of a weakly oscillated spectrum ($0.2\%$ deficit) at the closest location and a maximally-$\thone$-oscillated spectrum at the farthest location.
The $1.3$~km distance is optimal for the statistically-limited regime of a low power reactor.
Small flux monitor detector(s) benefit from the overburden inside the cavern and can calibrate the source normalization across the two locations.
\item[] \textbf{Configuration~B}: a $\lesssim 400\mwth$ reactor placed on the surface, at $0.7$ and $2.1\,\km$ from the detector.
Given the surface placement, the reactor power can be higher than in Configuration~A.
We find that the further far-site $\simeq 2$~km location is optimal for two reasons: i) this configuration is not limited by statistics due to the higher power reactor allowing us to slide the $\nuebar$ survival minimum to higher energies, and ii) no location on the surface would see an unoscillated spectrum, so having a further far-site location maximizes the difference between the near and far reactor $\nuebar$ spectrum.  
Small monitor detector(s) that travel with the reactor can help calibrate the emitted $\nuebar$ flux across stops but would require some overburden to reject cosmic backgrounds.
\end{enumerate}

In a companion paper, we explore other configurations for a fixed compact nuclear reactor source at JUNO without focusing on three-flavor oscillations~\cite{HMBP_inprep}.
The logistics of fixed microreactors coupled to large underground detectors is simpler and enables a high-statistics IBD and neutrino-electron scattering program.
In this sense, the oscillation program discussed here is more ambitious and imposes more constraints on the setup than other complementary programs on precision or beyond-the-Standard Model physics with fixed microreactors would.

It is worth noting that the idea of studying neutrinos with a movable reactor is not new: nuclear-powered vessels~\cite{Detwiler:2002ym,Cogswell:2021lla}, dedicated low-power reactors next to a large detector~\cite{Kornoukhov:2003ye}, and a mobile sub-$\!\gwth$ source operated at two distances from a $50\,$kt scintillator detector~\cite{Kopp:2006mw} have been previously discussed. 
Movable detectors have a longer history still: the same detector was operated at several distances at G\"{o}sgen, Bugey and Rovno in the 1980s~\cite{CALTECH-SIN-TUM:1986xvg,Cavaignac:1984sp,Afonin:1987gi,Declais:1994su}, movable detectors were proposed for cross-calibrating the near and far sites of the $\thone$ experiments~\cite{Shaevitz:2003ws,Anderson:2004pk,Bolton:2005yd,DayaBay:2007fgu}, a movable spectrometer scanning $50 - 1700\,$m from a reactor was proposed for sterile-neutrino searches~\cite{Kopeikin:2003uu}, and the same-detector ratio method is now standard at $\simeq 10$~m baselines~\cite{PROSPECT:2015iqr,DANSS:2018fnn,NEUTRINO-4:2018huq}. 
In this study, we consider a large detector already in operation, namely JUNO and its $20\,$kt of liquid scintillator, and a more quantitative treatment of two key systematics for these setups: the drift of the fuel composition between stops and the reproducibility of the emission rate between stops. 


This article is organized as follows.
We describe the oscillation signal in \cref{sec:oscillations} and the compact nuclear reactors and experimental configurations in \cref{sec:reactors}.
The sensitivity and statistical analysis are discussed in \cref{sec:analysis} and our results presented in \cref{sec:results}, including a discussion on the limitations and potential improvements to our approach.
We conclude in \cref{sec:conclusions} and leave the details of flux modeling and systematics to the appendices.



\section{Reactor antineutrino oscillations}
\label{sec:oscillations}

\begin{figure*}[t]
\centering
\includegraphics[width=0.9\columnwidth]{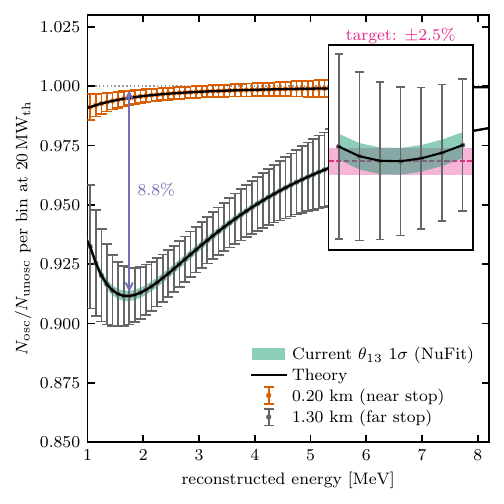}
\includegraphics[width=0.9\columnwidth]{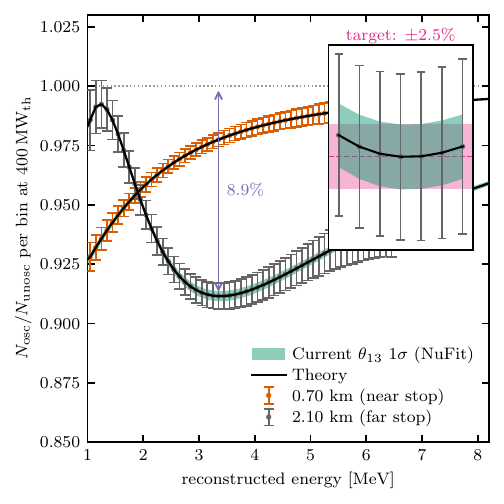}
\caption{
  Ratio of the oscillated to the unoscillated IBD spectrum, per reconstructed-energy bin of the JUNO 2025 release, at the two stops of each configuration, with the statistical uncertainties of the baseline programs.
  Left: Configuration~A, $20\mwth$, ten years split $25\% / 75\%$ between near and far. 
  Right: Configuration~B, $400\mwth$, four years split $30\% /70\%$ between near and far.  
  Orange: all near reactor runs combined; gray: the far reactor run.
  The bands show the change of the ratio for $\relsth = 2.5\%$, the present NuFit~6.0 precision. 
  In Configuration~B the deficits of the two stops sit at opposite ends of the spectrum, which makes their combined sensitivity to $\thone$ robust despite the lack of an unoscillated flux measurement.
  }
\label{fig:pee_stops}
\end{figure*}

At short baselines ($L\lesssim3\,\km$, $E\sim2 - 6\,\MeV$) the three-flavor survival probability reduces to the two-flavor form~\cite{Nunokawa:2005nx,Minakata:2006gq,Parke:2016joa} (exact expression in \cref{app:oscillations}),
\begin{align} \label{eq:approx}
  P_{ee}(E,L) &\simeq 1 - \sinsqtwo{\thone}\sin^2\left(\frac{\Delta m^2_{ee} L}{4E_\nu}\right),
\end{align}
with $\dmee$ as defined in \cref{eq:DayaBay_Gd}.
The first survival minimum (disappearance maximum) occurs at
\begin{equation}
  \frac{L_{\rm min}}{E} = \frac{\pi}{2\times1.267\,\dmee}
    \simeq 0.49\, \text{km}/\MeV.
\end{equation}
Therefore, $L_{\rm min} \sim 1.8\kilometers$ for the $\nuebar$ spectrum peak of $E \simeq 3.5 \MeV$. 
At this minimum, $P_{ee}^{\rm min} = 1-\sinsqtwo{\thone}\sim 0.913$, producing a $8.7\%$ dip controlled primarily by $\thone$. 
Resolving this dip at $n\sigma$ in a statistically-dominated measurement requires a minimum number of events of
\begin{equation}
  \label{eq:Nstat}
  N_{\rm min} \simeq
  \left(\frac{n}{\sinsqtwo{\thone}}\right)^{\!2}
  \simeq 140 \,n^2 \text{ events},
\end{equation}
giving $N_{\rm min}\simeq 560$ ($2\sigma$) or $1260$ ($3\sigma$) events. 
More generally, a fractional precision $\delta \equiv \sigma(\sinsqtwo{\thone})/\sinsqtwo{\thone}$ corresponds to $n = 1/\delta$, so $N_{\rm min} \simeq 140/\delta^2$.
For example, $\delta = 12\%$ for $10^4$ events, and Daya Bay's $2.8\%$~\cite{DayaBay:2022orm} for $\simeq 2\times10^5$.

The above argument assumes that all events sit at the dip and that the unoscillated normalization is known.
An uncorrelated normalization uncertainty $\sigma_f$ adds in quadrature,
\begin{equation}
  \label{eq:eps_total}
  \delta^{2} =\simeq
  \left(\frac{1-\sinsqtwo{\thone}}{\sinsqtwo{\thone}}\right)^{2}
  \left(\frac{1}{N} + \sigma_f^2\right),
\end{equation}
with $N$ the number of oscillated events at the dip.
A $2\%$ normalization uncertainty alone thus limits $\delta \gtrsim 21\%$ unless it cancels in a relative measurement.
Conversely, if a precision of $1\%$ on $\sinsqtwo{\thone}$ is desired, one must collect millions of events and control normalization systematic uncertainties to the level of $\sigma_f \lesssim 0.1\%$.
While only approximate, these numbers illustrate the key challenge behind a $\thone$ precision oscillation program. 

Normalization uncertainties common to all stops such as total flux, IBD cross section, and detection efficiency uncertainties cancel, and spectral uncertainties common to all stops are calibrated by the near reactor stop.
The dominant systematic is therefore the reproducibility of the $\nuebar$ rate across different reactor stops and of the detector response, including differences in fuel state and in detector volume and efficiency over time (\cref{sec:results}).
To mitigate this issue, a flux monitor detector that moves with the reactor can be considered.

With a flux monitor, the observable that drives the $\thone$ fit is, schematically, a double ratio
\begin{align}
\label{eq:double_ratio}
\mathcal{R}(E) &= \frac{\big(N_{\rm JUNO}(E)/N_{\rm monitor}(E)\big)_{\rm far}}{\big(N_{\rm JUNO}(E)/N_{\rm monitor}(E)\big)_{\rm near}}
\\\nonumber
&\simeq \left(\frac{L_{\rm near}}{L_{\rm far}}\right)^{2}\frac{P_{ee}(E,L_{\rm far})}{P_{ee}(E,L_{\rm near})},
\end{align}
where $L_{\rm near}$ and $L_{\rm far}$ correspond to the baseline of the near and far reactor runs.
In the double ratio, the reactor power, the yield per fission, and the fuel composition cancel together with any normalization common to all stops.
The main uncertainties that remain are the stability of both detectors across different reactor stops and the statistical uncertainty of the flux monitor detectors,
\begin{equation}
\label{eq:sigma_rel}
\sigma_{\rm rel}^2 \simeq \sum_{s\,\in\,\{\rm near,\,far\}}\left(\sigma^2_{{\rm JUNO},s}+\sigma^2_{{\rm mon},s}+\frac{1}{N_{{\rm mon},s}}\right),
\end{equation}
to be compared with the difference of the average deficits, 
\begin{equation}
    \Delta D \equiv \langle 1-P_{ee}\rangle_{\rm far}-\langle 1-P_{ee}\rangle_{\rm near}.
\end{equation}
In a rate-only measurement, we have that $\delta \simeq \sigma_{\rm rel}/\Delta D$, so for an average deficit of $\Delta D\simeq 5\%$, even a relatively small relative error $\sigma_{\rm rel} \simeq 0.1\%$ leads to a sizable cost in $\delta \simeq 2\%$.
This simple calculation highlights the main challenge of improving the precision on $\thone$.

For more on the interplay of correlated and uncorrelated normalization errors in multi-detector multi-reactor experiments, see Ref.~\cite{Kopp:2006mw} for mobile reactor and Ref.~\cite{Sugiyama:2004bv} for conventional multi-reactor setups.
Ref.~\cite{Kopp:2006mw}, in particular, considers the case of a moving active core, which bypasses many of the challenges presented by restarts of the fuel burnup, but at JUNO, this is likely only achievable if a new water channel were to be dug, in which a barge with an active reactor could be placed.

\Cref{fig:pee_stops} shows the ratio of the oscillated to the hypothetical unoscillated spectrum at JUNO for Configurations A and B.
The error bars indicate the statistical error on the event rate and the bands show the spread on the prediction due to the current uncertainty on $\thone$.
In the inset of each panel we show a zoomed-in view of the oscillation feature at the first survival minimum, where the oscillation amplitude is largest and the sensitivity to $\thone$ is maximized.
Note that while the individual bins carry statistical errors well beyond the shaded band, the target precision is reached through the combined fit of all bins at both stops.
In Configuration~B, the oscillation deficit at the near reactor site $L\simeq 0.7\kilometers$ is already $2.5\%$ and strongly energy dependent due to $\sin^2\Delta_{ee}$.
Nevertheless, this configuration provides a two-baseline measurement between a partially-oscillated pad and a fully-oscillated far-reactor configuration, together with the unoscillated spectrum measured by the flux monitor detector, effectively canceling systematics.

\section{Compact reactor sources for JUNO}
\label{sec:reactors}

A precision $\thone$ program needs an intense, compact and well-characterized $\nuebar$ source that can be operated at kilometer-scale distance from the detector. 
In addition to time of exposure, three key properties that determine the sensitivity of the measurement are the reactor thermal power, the dependence of the $\nuebar$ spectrum on the isotopic composition of the fuel and its evolution, and the reproducibility of the $\nuebar$ rate between run configurations (near site versus far site, for example).
With our benchmarks, the latter is likely to be the bottleneck for improving the precision on $\thone$.

\subsection{Fuel composition and the antineutrino spectrum}

A reactor emits about $2\times10^{20}$ $\nuebar$ per second per GW$_{\rm th}$, roughly six per fission, from the $\beta$ decays of the fission fragments of \uthirtyfive, \uthirtyeight, \pthirtynine and \pfortyone. 
In a given cycle, the antineutrino flux at some distance $L$ from the core (assumed to be point-like here) is given by
\begin{align}
\label{eq:source}
\frac{d\Phi_\nu}{dE_\nu}(E_\nu,\beta)&=\frac{P}{4\pi L^2}\frac{\sum_i f_i(\beta) S_i(E_\nu)}{\sum_j f_j(\beta)\,E_j},
\end{align}
with $S_i$ the per-fission antineutrino spectra, $E_i$ the average thermal energies per fission for a given isotope~\cite{Ma:2012bm}
\begin{equation}\label{eq:fission_energies}
E_i=(202.36,\,205.99,\,211.12,\,214.26)~\text{MeV},
\end{equation}
and $f_i$ the fission fraction per isotope $i$ ($i=\,^{235}$U, $^{238}$U, $^{239}$Pu, $^{241}$Pu; $\sum_if_i=1$).
As the fuel burns, the fraction of neutrino events from different isotopes changes, and since the four isotopes emit slightly different $\nuebar$ spectra~\cite{DayaBay:2019yxq,DayaBay:2021dqj,DayaBay:2025ngb}, the $\nuebar$ energy spectrum also changes, typically becoming lower energy (see \cref{app:flux}).

The fission fractions are set by the enrichment and the neutron spectrum, and evolve mainly through \pthirtynine breeding from \uthirtyeight, as directly observed by Daya Bay~\cite{DayaBay:2017jkb,DayaBay:2025ngb} and RENO~\cite{RENO:2018pwo}.
In a commercial low-enriched uranium (LEU, ${\lesssim}5\%$ \uthirtyfive) core, the \pthirtynine share grows from $\simeq 23\%$ to $\simeq 36\%$ over an 18-month cycle~\cite{DayaBay:2017jkb}. 
Since \pthirtynine yields $\simeq 30\%$ fewer IBD events per fission than \uthirtyfive~\cite{DayaBay:2019yxq}, the rate per unit power falls by several percent, and the high-energy spectrum by up to $\mathcal{O}(10\%)$.
Highly enriched (HEU, $>20\%$) research reactors are instead dominated by \uthirtyfive, which is why its spectrum is the best measured~\cite{PROSPECT:2018snc,Stereo:2021wfd,DayaBay:2021dqj,Kopeikin:2021ugh,DayaBay:2021owf}, while that of \uthirtyeight is the least~\cite{Haag:2013raa}.

Many small modular reactor (SMR) and microreactor designs use high-assay low-enriched uranium (HALEU, $5$–$20\%$ \uthirtyfive) to raise the power density and extend the core lifetime without refueling.
These often use tristructural isotropic (TRISO) fuel made of small uranium kernels coated in three layers of carbon-based materials that are designed to contain the fuel and fission products even under extreme temperatures.
For our purposes, the key characteristic of HALEU cores is the slow fuel evolution.
For the $20\mwth$, $19.75\%$-enriched thermal core of \cref{app:flux}, about $3\%$ of the \uthirtyfive inventory is burnt per year, the \pthirtynine fission fraction grows by $\simeq 0.5\%$ per year, and \uthirtyfive produces $>93\%$ of the $\nuebar$ flux during the entire program.
This has a small impact on the measurement because the IBD rate per unit power falls only by $\simeq 0.15\%$ per year, and the flux near $8\MeV$ by $\simeq 3\%$ in ten years, compared with $\simeq 12\%$ per 18-month LEU cycle.

If the source is switched off and on again, such as during a period of relocation, the low-energy spectrum ($E_\nu\lesssim 3.5\,\MeV$) requires days to weeks to re-equilibrate because of transient isotopes with half-lives of days to weeks, such as ${}^{140}$Ba ($t_{1/2} = 12.75$~days) and ${}^{140}$La ($t_{1/2} = 1.68$~days) (see, e.g., the recent measurement of spent nuclear fuel by Double Chooz in~\cite{DoubleChooz:2025jal}).
Longer-lived products like ${}^{106}$Ru ($t_{1/2} = 372$~days) and ${}^{144}$Ce ($t_{1/2} = 285$~days) build up over the first year of a fresh core and decay to very low energy neutrinos.
However, some of their $\beta^-$ daughters, such as $\isotope[144]{Pr}$ ($t_{1/2} = 17$~mins) and $\isotope[106]{Rh}$ ($t_{1/2} = 30$~s), can decay into neutrinos with $\lesssim 3.5$~MeV in energy.
Therefore, the fuel state at different stops differs, especially at low energies, introducing an important systematic uncertainty for precision oscillation studies with movable reactors.
This is the primary reason why we consider small or mid-sized ``flux monitor" detectors to calibrate the spectrum.

\subsection{Configuration A: underground microreactors}

JUNO's experimental hall lies $\simeq 690$~m below the surface and is reached by a $\simeq 1.3$~km long access tunnel with a $\simeq 42\%$ grade and a $5.7 \times 5.6$~m$^2$ cross section, and is accessible by a rail car and by a $564\,$m vertical shaft~\cite{JUNO:2015sjr,JUNO:2021vlw}. 
We estimate that the tunnel points approximately towards the JUNO detector, such that the distance between the center of the JUNO detector and a point in the access tunnel grows as $L \simeq 100\text{ m}+s$ with the distance $s$ along the tunnel from its bottom.
The tunnel spans exactly the baselines of interest, from ${\sim}0.2\,\km$ just above the hall to ${\sim}1.3\,\km$ close to the surface entry point, with the source always approximately $20^\circ$ above the detector. 
The cable train is able to transport a container as large as $4 \times 3 \times 12 \cubicmeters$~\cite{JUNO:2023ete}. 
Presumably, the incline is the laboratory's evacuation route, so a parked unit(s) would require an excavated alcove(s) to the side of the tunnel, potentially sealed from it.

The power of a reactor that can be operated in an active underground laboratory is limited by heat rejection and radiation levels.
We address the former first. 
Typical modern compact reactor designs under development are air-cooled and reject their heat to the ambient air. 
Underground, rejecting $1\,$MW into air with only a $10^\circ$~C temperature rise requires $\simeq 324,000$~m$^3$/h (for an air specific heat capacity of $c_p = 1$~kJ/kg.K), which is twice the laboratory's entire ventilation rate of $\simeq 160,000$~m$^3$/h~\cite{Cui:2023fcd}.
Dedicated exhaust ducts or a shaft that leads to the alcoves could in principle provide additional cooling.
We note that it is standard practice for deep mines to operate a ``surface-chilled water loop'' with surface ``chillers" or cooling towers that circulates water between the surface and the mine.
Such strategies would likely be a requirement for a $10 - 20\mwth$ source underground.
Any such system would need a dedicated feasibility study that also investigates potential injection of radioactive backgrounds throughout the laboratory, such as $\isotope[222]{Rn}$.

Underground reactors have been built and operated in several countries over the past decades, although these tend to be less than $250$~m deep and in large purpose-built rock caverns.
For example, Chooz A was an underground $\sim1\gwth$ reactor sited $150$~m under the surface, no longer operational since 1991. 
In Krasnoyarsk, an underground reactor site with 600~m.w.e. was considered for a neutrino oscillation program, Kr2Det~\cite{Martemyanov:2002td}, and is no longer operational.
A different approach has recently been discussed in the private sector in the United States.
Deep Fission Inc. is actively developing a deep borehole pressurized water reactor, with a single unit designed to deliver $45\mwth$ power a mile underground~\cite{deepfission2026gravity} (see also \cite{RAMIREZ202617}).
The reactor would operate under the pressure of the water column above it at a depth that exceeds that of the JUNO hall.
In principle, such mile-deep reactors can serve as fixed far reactors to JUNO, but calibrating their flux might require deep detectors, possibly in neighboring boreholes.

Radiation is unlikely to be a concern for JUNO itself since the rock provides natural shielding from the alcoves. Neutrino experiments routinely operate $10$–$50\,$m from $60\mwth$–$4.6\gwth$ cores, far closer than our minimum of $200\,$m.
Avoiding backgrounds to JUNO's low-energy single-hit program (e.g., solar neutrinos) will be a stricter requirement than meeting personnel-dose limits, although the JUNO veto may help.
The nearby flux monitors are more exposed and would need standard hydrogen-rich shielding (borated or Li-doped polyethylene) with $\mathcal{O}(10)\,$cm of lead~\cite{PROSPECT:2018dtt,STEREO:2018rfh,JUNO:2020ijm}, or a separate alcove.

Given the above considerations, we adopt a single low-power $\mathcal{O}(5 - 20\mwth)$ reactor for Configuration~A. 
We adopt $20\mwth$ as an optimistic benchmark, with a $5$~t flux monitor detector $10$~m from the core either in the same or in an adjacent alcove. 
Movable microreactors of this sort are being actively developed as part of the Department of Energy (DOE) Reactor Pilot Program for advanced reactor demonstration.
They typically consist of units of $\mathcal{O}(1 - 5 \mwth)$ power in one or a few shipping containers, designed to be deployed and then removed after operation.
Most projects we are aware of plan to use HALEU TRISO fuel to compensate for the smaller core and increase power density and overall efficiency, although some recent designs still consider LEU light-water reactor cores. 
For concreteness, we adopt the HALEU fuel for Configuration~A.

As of this writing, a number of microreactor projects have reached zero-power criticality, but none has operated at power or been relocated with irradiated fuel, so, to the best of our knowledge, the technology of a land-based movable reactor is yet to be demonstrated.
Each relocation would presumably involve a shutdown, disassembly, the transport of the core with or without the spent fuel, and a restart period~\cite{PNNL:MicroreactorTransport}. 
While not a limitation in principle, this makes the reproducibility of the emission rate between stops a key factor to be considered.
Finally, we note that since power generation may not be the primary goal of these microreactors, their operational profiles could be quite different from existing projects.
Given the uncertainties involved, our choices should be viewed as exploratory and not final.

\begin{figure*}[t]
\centering
\includegraphics[width=\textwidth]{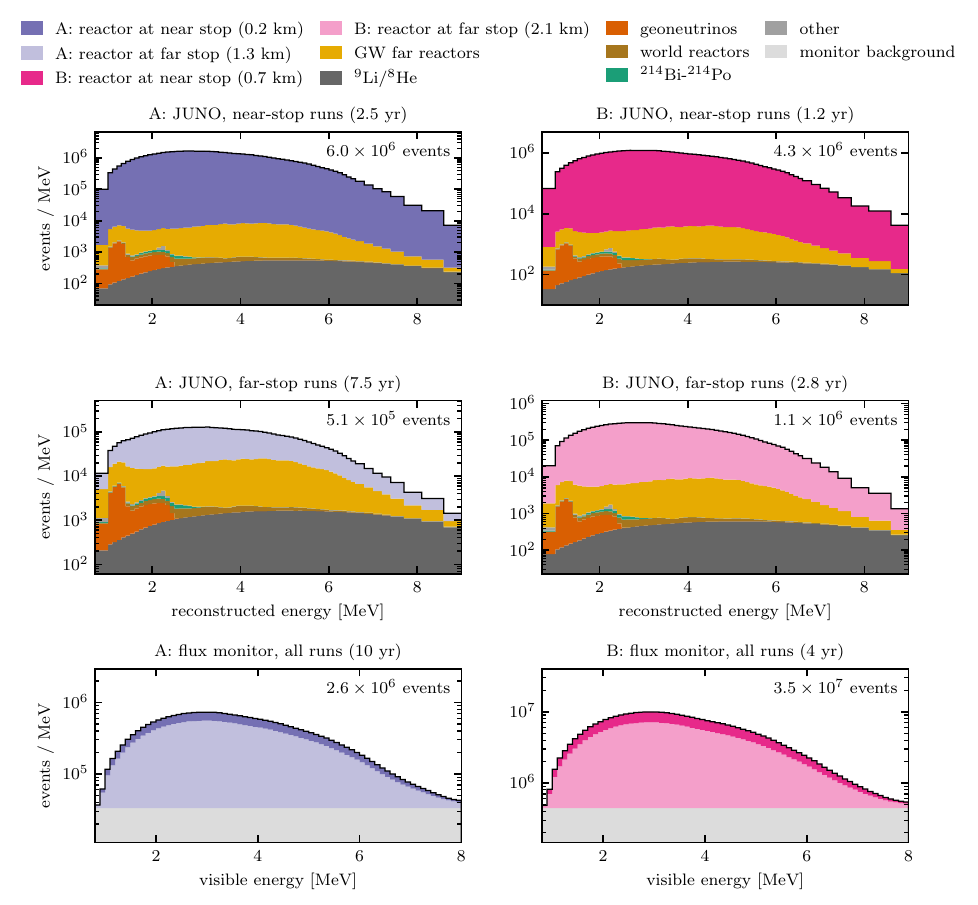}
\caption{
  The event rates used in our analysis for Configuration A (left) and B (right) for the JUNO detector with the compact near reactor rates at the near location (top row) and far location (middle row). 
  We also show the rates for the flux monitor detector (bottom row).
  The backgrounds in the JUNO detector are taken from the data release in \cite{JUNO:2025gmd} while for the flux monitor detectors we assume a constant rate of $\simeq 660$ IBD/day in scenario A and a starting rate of $\simeq 2\times10^4$ IBD/day for scenario B (see text).
  }
\label{fig:event_rates}
\end{figure*}

\subsection{Configuration B: surface reactors}

Powers of $\mathcal{O}(100\mwth)$ and above exist only as fixed small modular reactors (SMRs), such as the $385\mwth$ ACP-100 in China~\cite{SMR_ACP100}, or as floating units, such as the two $150\mwth$ KLT-40S cores of the \emph{Akademik Lomonosov} floating nuclear power station~\cite{AkademikLomonosov} and the $175/200\mwth$ RITM-200 series of icebreakers~\cite{WNA:NuclearShips,IAEA:SMR2022}. 
An SMR of this size can only realistically be placed on the surface.\footnote{To our knowledge, no navigable waterway reaches the JUNO laboratory site. 
If such waterway existed or could be built, then a continuously-operating core would be far more advantageous for oscillation physics, providing a full picture of the oscillogram plane of $L$ vs $E_\nu$.
} 
The closest distance between the JUNO detector and the surface is about $0.66\kilometers$ away at the shaft collar.
The hill top is about $0.69\kilometers$ from the center of the JUNO detector, and the tunnel entry is about $\gtrsim 1.3\kilometers$ away~\cite{JUNO:2015sjr,JUNO:2021vlw,JUNO:2023ete}.
In addition to the higher power involved, a key difference of placing reactors on the surface is that no location allows JUNO to measure an unoscillated spectrum.
At $0.7\kilometers$ away the $\nuebar$ deficit is already $1 - \langle P_{ee}\rangle \lesssim 3\%$, so a measurement of $\thone$ is not a simple ratio of oscillated to unoscillated spectra.

In this configuration, we consider instead two ``surface pads".
One at $0.7\,\km$ and another at $2.1\kilometers$ away from the center of the JUNO detector, beyond the first oscillation minimum of the low-energy neutrinos (\cref{sec:oscillations}).
As before, to help mitigate any residual flux systematics we add a small flux monitor detector that follows the reactor wherever it is deployed, without necessarily trying to cancel detector-related systematics with the main JUNO detector.
This technique only works if the detector systematics of the flux monitor is canceled or well below the statistical uncertainty of the measurement between different surface pads, which seems more feasible than for the reactor flux itself.
Note that compact, transportable antineutrino detectors have already been operated at reactors for monitoring and safeguards~\cite{Bernstein:2019hix,Haghighat:2018mve,Sutanto:2021xpo,Shi:2026spq}.

Finally, we note that \emph{replica} cores of identical design and fuel operated in sequence at fixed baselines are an alternative to relocating the reactor, both in Configuration~A and Configuration~B.
By using replicas, one would trade the logistics of moving the reactor for that of carefully manufacturing and operating two cores identically at two different periods of time so as to maintain the $\nuebar$ spectrum the same from core to core and over time.
The flux monitor detector is also crucial in this case, as well as the control of detector systematics over time.

\subsection{Flux Monitor Detectors}

As illustrated by \cref{eq:double_ratio}, the presence of flux monitor detectors can cancel flux changes across runs and is critical to the success of a mobile reactor.
These monitors do not need the resolution or the background suppression of JUNO or JUNO-TAO because the $\thone$ dip is broad in $L/E$, so calibrating narrow or small features in the spectrum does not significantly improve sensitivity to $\thone$.
For concreteness, we consider a $5$~t ($20$~t) Gd-loaded scintillator detector at $10$~m ($15$~m) from the core in Configuration A (B) to measure the $\nuebar$ emission rate as accurately as possible, beating the dominant cosmogenic backgrounds.
At $400\mwth$, the Configuration B flux monitor records $\simeq 2\times10^4$ IBD events per day ($\simeq 2\times 10^3$ background events), enough to fix the rate ratio between pads to $0.05\%$ statistically within each stop and to track the fuel evolution of an LEU core. We predict that the fuel evolution will cause a drop of $\sim 2.5\%$ in expected events mid-cycle (0.75 years) and $\sim 5\%$ by the end of one fuel cycle (1.5 years).
With shielding for such a monitor detector of about $\gtrsim 100$~m.w.e of overburden (e.g., a $\simeq 40\,$m pit beside each pad), the signal-to-background ratio can reach above ten~\cite{DoubleChooz:2019qbj,NEOS:2016wee,DANSS:2018fnn}.


\subsection{Schedules and Fuel Evolution}

Finding an optimal schedule of the near and far-site reactor stops is also important.
Dividing the stops evenly in time is generally a bad strategy for two reasons: the far stop requires longer exposure to accumulate sufficient statistics, and the fuel evolution probed during the near stop may be significantly different from that during the far stop.
The \uthirtyfive$\to$\pthirtynine evolution (see \cref{app:flux}) leads to a smooth tilt of the $\nuebar$ spectrum that is degenerate with the depth of the oscillation dip. 
Splitting the near-site stop into two halves before and after the far-site stop can help mitigate the degeneracy between fuel evolution and the oscillation dip.
This bracketing strategy is used in all of our configurations.
The exact ratio between near-site and far-site exposure times depends on the specific reactor configuration and the desired precision on $\thone$, and is numerically optimized for each scenario.

The near reactor fluxes are predicted as in \cref{app:flux}.
For Configuration A, the flux is calculated assuming a HALEU microreactor with a starting fuel composition of $f_{235}=0.98$ and $f_{238}=0.02$, evolving the fuel composition over time according to the reactor burnup.
For Configuration B, we assume a conventional LEU core following the fission fractions measured by Daya Bay over an 18-month cycle~\cite{DayaBay:2017jkb}.
In both cases, the spectrum is determined by the Daya Bay \uthirtyfive and \pthirtynine yields for the compact source, with ${}^{238}$U and \pfortyone from the Huber-Mueller model~\cite{Huber:2011wv,Mueller:2011nm}.

\Cref{fig:event_rates} shows the event rates for the JUNO detector for near and far siting of the mobile reactors as well as of the flux monitor.
For JUNO, we use the background components provided by their data release~\cite{JUNO:2025gmd}.
For the monitor detector, we assume a constant 10\% background of 66 events/day (scenario A) and $2\times 10^3$ (scenario B) evenly spread across energy.


\section{Analysis method}
\label{sec:analysis}

\begin{figure*}[t]
  \centering
  \includegraphics[width=\columnwidth]{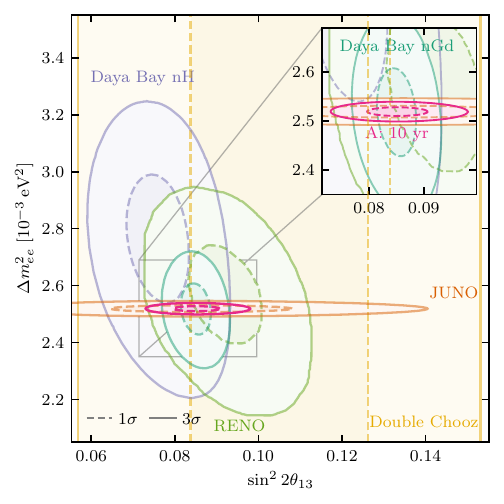}
  \includegraphics[width=\columnwidth]{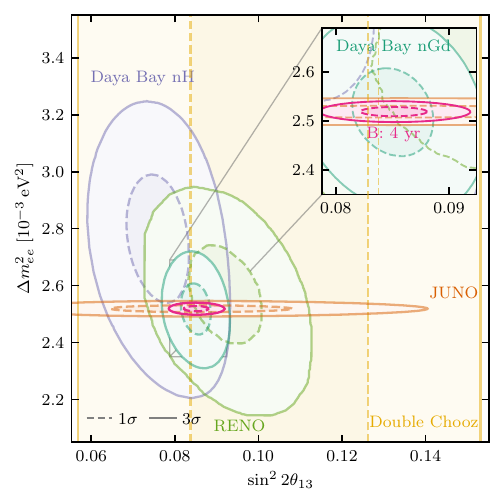}
  \caption{%
    Confidence regions ($1\sigma$ dashed, $3\sigma$ solid, two d.o.f.) in the $\dmee$ vs $\sinsqtwo{\thone}$ plane for JUNO's far reactors alone after six years (orange), for the near reactor program discussed here (pink) in Configuration A after 10 years (left) and B after 4 years (right).
    Both configurations include six years of standalone JUNO data in the fit.
    We assume an uncertainty on the reproducibility of the reactor flux between stops of $\sigma_P=0.5\%$.
    The final results from Daya Bay using only neutron capture on Gd~\cite{DayaBay:2022orm} or only capture on hydrogen~\cite{DayaBay:2024hrv} are shown as teal and blue, respectively.
    Also shown is the result from RENO~\cite{RENO:2024msr} (green).
    The Double Chooz~\cite{DoubleChooz:2019qbj} measurement of $\thone$ is shown as a vertical band as it assumes an external prior of $\Delta m^2_{ee} = (2.484 \pm 0.036) \times 10^{-3}$~eV$^2$.
      \label{fig:combined_200m}
    }
\end{figure*}

In this section we detail our analysis and statistical methodology.
There are many sources of uncertainty in our setup across detectors, reactors, and time periods, so we leave the details to \cref{app:flux,app:validation}.

We start with the IBD spectrum of each reactor run at JUNO and the flux monitor detectors.
These are calculated using the IBD cross section~\cite{Vogel:1999zy,Strumia:2003zx} and the Daya Bay total flux model applied to JUNO's GW-scale power plant cores and the Daya Bay \uthirtyfive and \pthirtynine yields for the compact source, with ${}^{238}$U and \pfortyone from Huber-Mueller.
At the far site of Configuration~A with the $20\mwth$ benchmark reactor the JUNO detector would see about $148$ IBD events per day at $1.3\,\km$, $4.2$ times the $\gwth$ far-reactor rate.
At the far site of Configuration~B, the $400\mwth$ mobile reactor at $2.1\,\km$ yields $1.1\times 10^3$ IBD events per day at JUNO, $30$ times the observed (oscillated) $\gwth$ far-reactor rate.
Therefore, the $\gwth$ far-reactor contribution, although subdominant, must be included in the analysis.

We validate the far-reactor model by approximately reproducing the standard JUNO $\theta_{12}$ vs $\Delta m^{2}_{21}$ measurement, and we find good agreement with the published measurements (see \cref{app:validation}).
GW-scale power plants include the oscillated spectrum of the two $4.6\gwth$ Taishan reactors at $L = 52.5\,\km$, the six $2.9\gwth$ Yangjiang reactors at $L = 52.5\,\km$, and the six $2.9\gwth$ Daya Bay reactors, treated as one effective core at $L\simeq 215\,\km$.
This far-reactor prediction is normalized to JUNO's published non-oscillated and efficiency-corrected rate of $150.9$~events/day~\cite{JUNO:2025gmd}, which accounts for the reactor live-times, multiplied by the $69.9\%$ selection efficiency. 
This gives $35.1$ oscillated events per day.
We then rescale this result for different exposure times as needed.

To model the JUNO detector energy resolution, we use JUNO's released positron nonlinearity and energy resolution.
The energy response function is modeled as follows:
\begin{align}
\label{eq:response}
\tilde E_{\rm pr} &= E_{\rm pr}\,r_{\rm nl}\big[(1+\xi_{\rm scl})F_{\rm nl}(E_{\rm pr})+\xi_{\rm bias}\big],
\end{align}
where $E_{\rm pr}$ is the true positron energy, $\tilde E_{\rm pr}$ is the reconstructed positron energy, $r_{\rm nl}$ is the non-linearity scaling factor, $\xi_{\rm scl}$, $\xi_{\rm bias}$, and $\xi_{\rm res}$ are the corresponding nuisance parameters, $F_{\rm nl}$ is JUNO's positron non-linearity from \cite{JUNO:2025gmd}.
The corresponding energy resolution is given by
\begin{align}
\sigma(\tilde E) &= (1+\xi_{\rm res})\,r_{\rm res}\,\sqrt{a^2\tilde E+b^2\tilde E^2+c^2},
\end{align}
where $r_{\rm res}$ is the resolution scaling factor and $a=3.18\%$, $b=0.95\%$, and $c=0.46\%$ are fitted to the eight calibration peaks of the 2025 data release~\cite{JUNO:2025gmd}.
The binning follows the 2025 data release~\cite{JUNO:2025gmd}: 66 bins from 0.7~MeV to 12~MeV.
For the small flux monitor detector we use an $8\%/\sqrt{E}$ resolution with $0.1\,\MeV$ bins.
We have checked that using a smaller resolution, like $2\%/\sqrt{E}$ as in JUNO-TAO, does not significantly impact our sensitivity to $\thone$.

Our statistical sensitivity is quantified with the Asimov $\chi^2$,
\begin{align}\label{eq:asimov_chi2}
  \chi^2(\thone,\dmee) = \mathbf{r}^T C^{-1} \mathbf{r},
\end{align}
with $C$ the joint covariance of all samples across reactors, detectors, and run periods, and
\begin{equation}\label{eq:residuals}
  \mathbf{r} = \mathbf{N}_{\mathrm{obs}} - \mathbf{N}_{\mathrm{pred}}(\thone,\dmee)
\end{equation}
the vector of residuals calculated for $\mathbf{N}_{\rm obs}$ evaluated using the oscillation best-fit values from NuFit~6.0~\cite{Esteban:2024eli}.
The construction of the covariance is explained in detail in \cref{app:flux}, but in general it takes the form
\begin{equation}
C = \mathrm{diag}(\boldsymbol{\mu}) + \sum_k\mathbf{v}_k\mathbf{v}_k^T,
\end{equation}
where $\boldsymbol{\mu}$ are the predicted counts of all samples for a given exposure, and $\mathbf{v}_k = \sigma_k \, \partial \mathbf{N}/\partial \eta_k$ 
is the shift of the counts for a one-sigma change of the nuisance parameter $\eta_k$.
The covariance includes the $75\times75$ Daya Bay per-isotope flux covariance, the 1.6\% error on the JUNO efficiency, energy-scale, bias and resolution uncertainties shared with the far-reactor data, the released background components and their uncertainties. The fuel evolution of \cref{app:depletion} enters with a single, fully correlated $30\%$ uncertainty on the size of the burnup-driven change in the fission fractions.
Confidence regions at $1,2,3\sigma$ correspond to $\Delta\chi^2 = 2.30,\,6.18,\,11.83$ (two d.o.f.), and we profile for the one-dimensional uncertainties.

\begin{figure}[t]
  \centering
  \includegraphics[width=0.49\textwidth]{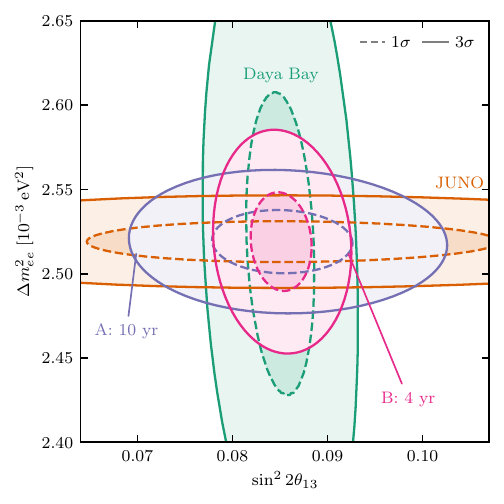}
  \caption{%
    A zoomed version of \cref{fig:combined_200m} including the sensitivity of Configuration A and B without standalone JUNO data.
    \label{fig:sensitivity_noJUNO}
    }
\end{figure}

Among the many systematic uncertainties we consider, the ones specific to our movable sources include the flux normalization at each reactor stop and the additional samples from the flux monitor detectors. 
We parameterize the power at each stop as
\begin{equation}
P_k = P\,(1+\epsilon)(1+\delta_k),
\end{equation}
with $P$ the nominal reactor power, $\epsilon$ a nuisance parameter with $2\%$ uncertainty on the absolute power that is common to all reactor locations and runs, and $\delta_k$ a nuisance parameter describing differences between locations and runs, unique to a given reactor at a given location and run with prior $\sigma_P$.
While $\epsilon$ affects all stops equally and is effectively constrained by the combined data, $\delta_k$ and its uncertainty $\sigma_P$ capture variations between stops and limit the reproducibility of the $\nuebar$ rate between stops.
As we will see, constraining $\sigma_P$ is crucial for the success of a movable-source program.
This would not be a concern if the reactor was moved continuously and never shut down between stops (as in the floating reactor concepts explored in \cite{Kopp:2006mw}), however, a land-based unit is unlikely to be operated in such a manner.

A key point of our setup is that the detectors are far more stable between stops than the reactor power itself.
Based on the experience from the Daya Bay detectors~\cite{DayaBay:2012aa,DayaBay:2022orm}, we add an independent $0.1\%$ uncertainty per-stop to parameterize changes to the detector efficiency and geometry between stops over time for both the JUNO and the monitor detectors.
As we will see, in the large statistics limit (achieved in Configuration~B), this uncertainty on the detector stability between stops will be one of the dominant bottlenecks on the precision on $\thone$.
Note that the flux monitor detectors could be either moving with the reactor or remain fixed if there is one at each reactor location.

The movable near reactor schedules are chosen so that they bracket the far-site stop with near-site stops before and after.
The fuel composition of the movable source is integrated over each stop in $60$~d steps. 
For Configuration~B we use two such near-far-near cycles to shorten each run duration.
This allows JUNO to measure the composition of the LEU core more evenly given a shorter burnup period with refueling every 18 months.
For each refueling, it may be advantageous to run the full cycle again in order to improve the reproducibility of the flux throughout the run.
The $\gwth$ far reactors contribute to every sample with a common set of systematic uncertainties and are described by the Daya Bay data release using the ``Total Spectrum" provided~\cite{DayaBay:2025ngb}.
Finally, note that fuel evolution is far slower for the $20 \mwth$ HALEU core considered here (see \cref{app:depletion}).


\section{Results and Discussion}
\label{sec:results}

We now present the results for the two configurations of \cref{sec:reactors}.

\subsection{Sensitivity to Oscillation Parameters}

\Cref{fig:combined_200m} shows the resulting confidence regions in the $\thone$ vs $\dmee$ plane for Configuration~A and B separately, as well as the standard JUNO $\gwth$ far reactor complex only measurement.
The latter constrains $\dmee$ beyond current global fit uncertainties, but does not have a competitive sensitivity to $\thone$.
\Cref{fig:sensitivity_noJUNO} shows a zoomed in version comparing only JUNO and Daya Bay.
In this panel we also show the results of the fit using only the compact mobile reactor dataset, showing how the sensitivity to $\dmee$ is somewhat weaker than standalone JUNO.
In Configuration~A and B, the near-reactor measurements significantly improve the sensitivity to $\thone$, reducing the JUNO far-reactor-only uncertainty on $\sinsqtwo{\thone}$ after six years by factors of 3.9 and 7.4, respectively, while also providing complementary sensitivity to $\dmee$.

Note that the method behind the $\dmee$ measurement is completely different at short baselines than at the long baselines of JUNO.
In the former, it is extracted from the first $\nuebar$ survival minimum, while in the latter, it comes from the interference pattern over multiple oscillation cycles at the $\Delta m^2_{21}$ minimum.
This complementarity would be a powerful tool to constrain any additional new physics affecting either short or long-baseline neutrino oscillations.

\begin{figure*}[t]
  \centering
  \includegraphics[width=\textwidth]{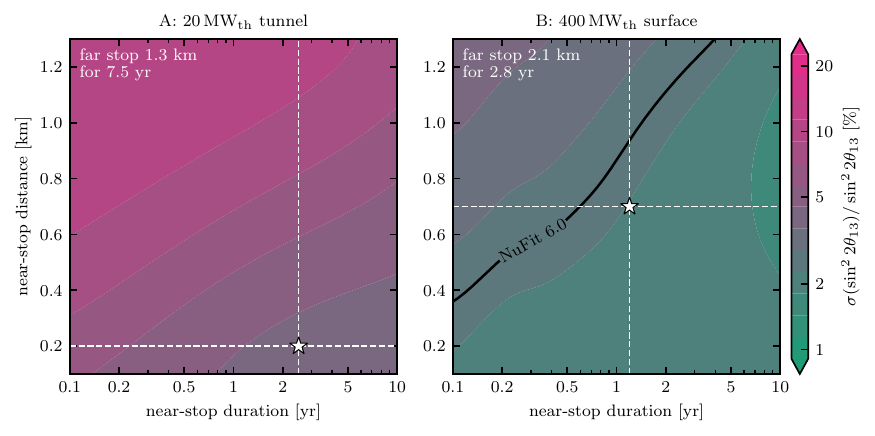}
  \caption{
  The achieved relative uncertainty on $\sinsqtwo{\thone}$ (color) for Configuration A (left) and B (right) as a function of the near reactor distance to JUNO and the duration of the near-reactor program.
  The black contour shows the NuFit~6.0 precision, where reached. Everywhere on these maps the precision is better than that of JUNO alone after six years ($16.6\%$). The far stop is held at its benchmark distance and duration.
  The white star shows the benchmark numbers used in \cref{fig:combined_200m}.
  }
  \label{fig:two_near_configs}
\end{figure*}

\begin{figure*}[t]
  \centering
  \includegraphics[width=\textwidth]{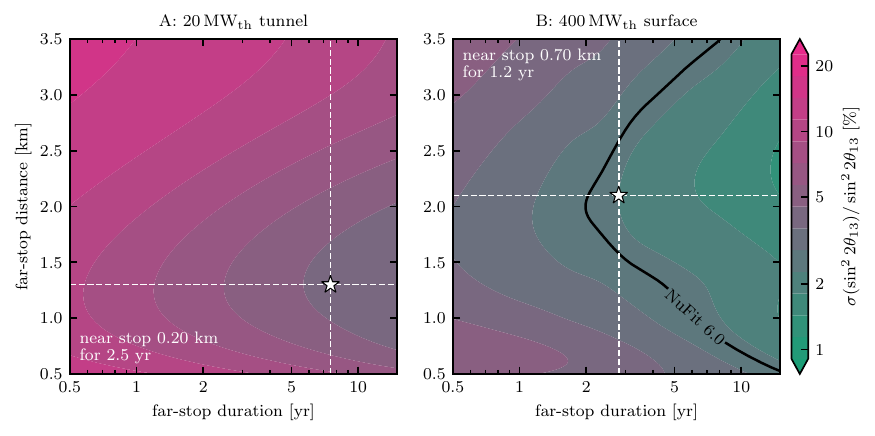}
  \caption{
  Same as \cref{fig:two_near_configs}, but varying the duration and baseline of the far reactor.
  }
  \label{fig:two_configs}
\end{figure*}

\Cref{fig:two_near_configs,fig:two_configs} show how the reach to $\thone$ evolves with the distance and duration of the near and far stops.
These scans use the default assumption of $\sigma_P = 0.5\%$ for the uncertainty on the power reproducibility between stops, with the flux-monitor detectors.
In \cref{tab:relnorm}, we compare this to the case without them and to the idealized case of $\sigma_P = 0\%$. 
It is evident that the flux-monitor detectors are crucial to mitigate the impact of the reproducibility uncertainties between stops.
Note that Configuration A setup, limited by the smaller reactor power, never really reaches the sensitivity of Configuration B.
Increasing the statistics, such as by increasing the microreactor power, would be the most important factor. 
This is not the case for Configuration B, where the higher reactor power allows to reach the Daya Bay precision in less than $3$~years.
Running for a decade would bring Configuration~B to $\relsth \simeq 1.6\%$. In all cases, the improvement on $\dmee$ is modest compared to what standard JUNO achieves.

\subsection{Impact of Systematics}

We now turn to the impact of the power reproducibility uncertainty between stops.
In our analysis, we find that the normalization errors that are common to all stops (e.g., $2\%$ absolute power and $1.6\%$ efficiency) have no effect on $\sigma(\sinsqtwo{\thone})$.
In fact, removing the Daya Bay flux covariance altogether barely changes $\sigma(\sinsqtwo{\thone})$, indicating the flux is being well calibrated.

In Configuration~A, the flux monitor detectors are statistics-limited and, therefore, cannot fully mitigate the effect of $\sigma_P$.
From $\sigma_P=0$ to $2\%$, the ten-year relative uncertainty on $\sinsqtwo{\thone}$ grows from $3.5\%$ to $4.3\%$ with the monitor and to $8.3\%$ without it. 
In Configuration~B, on the other hand, the flux monitor detectors remove the dependence on $\sigma_P$ significantly.
With the monitor, the four-year precision stays between $2.1\%$ and $2.3\%$ for any $\sigma_P$, while without it the precision degrades to $2.9\%$ and $3.0\%$ for $\sigma_P=0.5\%$ and $2\%$ (\cref{tab:relnorm}).
With the monitor, the limiting factor on the uncertainty is the detector stability uncertainty assumed to be $0.1\%$ here.
At first this may seem like a surprisingly stringent criterion, however, a $0.1\%$ on a deficit difference of $5\%$ translates to a $2\%$ error on $\sinsqtwo{\thone}$. 
We find that setting either the JUNO or the flux monitor detector normalization stability uncertainty to zero improves the precision on $\sinsqtwo{\thone}$ from $2.1\%$ to $1.9\%$ in 5 years.

Maintaining this level of reproducibility of the rate between stops requires both a stable detector as well as a stable reactor.
The former is not likely to be the limiting factor.
We note that Daya Bay's detectors were stable to ${\sim}0.1\%$ over years~\cite{DayaBay:2012aa,DayaBay:2022orm}.
Note that $0.1\%$ in rate corresponds to $10$\,cm in $L$ at $200$\,m. 
The hardest part will inevitably be the reactor. 
After a relocation the core is restarted and brought to a nominal power that commercial plants know to ${\sim}0.5\%$~\cite{DayaBay:2016ggj,Djurcic:2008ny}, and the reproducibility of a set point with the same instrumentation has not been characterized at the $0.1\%$ level for any reactor. 

To mitigate the reproducibility uncertainty, the flux monitor detector(s) are crucial.
For the $400\mwth$ reactor, the $2\times10^4$ events per day give $0.05\%$ statistical error per stop, meaning that only the flux monitor detector stability throughout runs enters as a significant uncertainty.
In Configuration~A, the flux monitor detector collects only $\sim 650$ events per day due to the smaller reactor power and constrains each reference block to $0.2\%$, so the program remains sensitive to $\sigma_P$ unless the monitor mass is increased.

Quantitatively, we find that in our analysis running the compact reactor at the near-site only at the beginning of the schedule (no bracketing) increases $\sigma(\sinsqtwo{\thone})$ by about $5\%$ compared to bracketing the near-site run around the far-site stop.
In practice, however, this leaves $\thone$ exposed to a bias from potential mismodeled fuel burnup.
We also find that a near-far-near bracketing schedule is sufficient and that finer interleaving of near and far reactor sites does not add much to the precision of $\thone$.
In particular, the low-energy spectrum needs days to weeks after a restart to return to equilibrium~\cite{Mueller:2011nm}, so every stop should begin a fixed interval after the restart, or the first weeks should be excluded from the fit.

Since the systematics that limit the measurement do not scale linearly with exposure, we quote our sensitivities for time durations with a fixed reactor power.
Increasing or decreasing the power of the reactor does not necessarily mean that the program length will decrease or increase linearly for the same sensitivity.


\renewcommand{\arraystretch}{1.3}
\begin{table}[t]
  \centering
  \begin{tabular}{c|cccc}
  \toprule
    \multirow{2}{*}{$\sigma_P$} & \multicolumn{2}{c}{A} & \multicolumn{2}{c}{B}
    \\
     & monitor & no monitor & monitor & no monitor 
    \\
    \midrule
    $0$      & $3.5\%$ & $3.5\%$ & $2.1\%$ & $2.2\%$ \\
    $0.1\%$  & $3.8\%$ & $3.9\%$ & $2.2\%$ & $2.4\%$ \\
    $0.5\%$  & $4.2\%$ & $6.6\%$ & $2.3\%$ & $2.9\%$ \\
    $2\%$    & $4.3\%$ & $8.3\%$ & $2.3\%$ & $3.0\%$ \\
    \bottomrule
  \end{tabular}
  \caption{Effect of the uncertainty on the power reproducibility per stop, $\sigma_P$, with and without the monitor detector, for JUNO.
  The uncertainty on the flux monitor detector(s) stability is fixed at $0.1\%$. Configuration~A ($20\mwth$, $0.20/1.3\,\km$): $\relsth$ after ten years. Configuration~B ($400\mwth$, $0.70/2.1\,\km$, two cycles): $\relsth$ after four years. Both include six years of standalone JUNO data.
  }
  \label{tab:relnorm}
\end{table}



Finally, we note that while the smooth spectral distortions from fuel evolution or from the reactor model are degenerate with the depth of the $\dmee$ dip, they are less relevant for rapid $L/E$ oscillations of an eV$^2$-scale sterile neutrino, for example.
At JUNO, its $35\,$m diameter can resolve these inside the detector as a clear oscillation pattern for a single reactor baseline.
Searches of this kind are insensitive to the flux model by construction, as demonstrated with segmented and movable detectors~\cite{PROSPECT:2018dtt,STEREO:2018rfh,DANSS:2018fnn}, and they, together with $\bar\nu_e$-electron scattering and searches for new physics and light new states, are another compelling use of a tunnel microreactor, even if its power falls short of the ambitious $\thone$ goal~\cite{HMBP_inprep}.


\section{Conclusions}
\label{sec:conclusions}
The technology of deployable compact reactors is actively being pursued worldwide.
These range from units of few $\mwth$ with typical dimensions of standard containers to small modular reactors of several hundred $\mwth$ that are assembled from prefabricated modules. 
These reactors will be intense deployable antineutrino sources that can be brought close to existing large detectors.
With this in mind, we consider a potential post-mass-ordering program for JUNO involving a deployable compact reactor.

JUNO will establish the mass ordering at $3\sigma$ in about six to eight years~\cite{JUNO:2024jaw}.
The final stretch of this period is likely to bring a greater-than-$5\sigma$ determination from global neutrino data including Hyper-Kamiokande and atmospheric neutrinos~\cite{Cabrera:2020ksc,Forero:2021lax,JUNO:2021vlw,JUNO:2024jaw}.
It could also coincide with the MO determination at more than $5\sigma$ at DUNE alone through matter effects~\cite{DUNE:2020lwj}.\footnote{For the current status of global constraints on the mass ordering and their caveats~\cite{Machado:2026shm}, see Refs.~\cite{JUNO:2026tap,NOvA:2026tir,Jin:2026ejo}.}
JUNO can then turn to other physics programs, including a possible upgrade to a neutrinoless double-beta decay experiment~\cite{Zhao:2016brs}.
Here we discussed another possibility: the deployment of a movable compact reactor either in the access tunnel of the underground laboratory or on the surface above it.
In this first exploratory study, we consider a measurement of $\thone$ and what would be needed to improve on the world-leading Daya Bay's measurement.


Using the Daya Bay \uthirtyfive and \pthirtynine spectra and covariance~\cite{DayaBay:2025ngb} and the Huber-Mueller model for \uthirtyeight and \pfortyone~\cite{Huber:2011wv,Mueller:2011nm}, we modeled a HALEU core with a simplified fuel evolution model and an LEU core following thr Daya Bay fuel evolution.
For the HALEU core, the high-energy flux component drops by only $\simeq 3\%$ in ten years, in contrast with an $\mathcal{O}(10\%)$ drop for thr 18-month cycle of the LEU core.
We find that Configuration A with a $20\mwth$ microreactor at $0.2$ and $1.3\,\km$ with a $5\,$t monitor, reaches $\relsth \simeq 5.3\%$ after five years and $4.2\%$ after ten.
This reduces to $3.5\%$ for a perfectly reproducible rate, still falling short of Daya Bay's $2.8\%$.
For Configuration B with a $400\mwth$ surface unit at $0.7$ and $2.1\,\km$, with a $20\,$t monitor, we find that JUNO could reach the present global-fit precision of $2.5\%$ in about three years. 
With more exposure, Configuration B could in principle meaningfully improve on Daya Bay's result.
The sensitivity of the near reactor datasets to $\dmee$ would be complementary to that of standalone JUNO, but it cannot improve upon it since the latter probes multiple oscillation cycles, including the interference with the solar mass splitting.

A key requirement for the success of the program proposed here is that the $\nuebar$ emission rate be reproducible, or monitored, to $\simeq 0.1\%$ between stops.
Smooth fuel-evolution distortions are degenerate with the depth of the broad $\thone$ dip (unlike, e.g., the rapid oscillations of eV-scale sterile neutrinos~\cite{PROSPECT:2018dtt,STEREO:2018rfh,DANSS:2018fnn}).
While part of this effect can be controlled by using a near-far-near reactor schedule that brackets the evolution, a flux monitor detector would most likely be critical.
Additional alcoves in Configuration~A, for example, could also provide more baselines and help separate fuel evolution from the $L/E$ dependence of oscillations.
Characterizing the stop-to-stop reproducibility of the emission rate and the achievable shape uncertainty should be the priority for future work in this direction.

Ideally, the reactor would be moving without being turned off.
This would be the case of a reactor mounted on a barge or floating units along a waterway or channel.
In this case, one can genuinely scan the baseline and construct an ``oscillogram" in $L$ and $E_\nu$, mapping the first and even the second $\thone$ oscillation dips.
We are not aware of any waterways around the JUNO location, but if such a channel could be built, a floating unit could deliver a substantial amount of power, such as the one considered in Configuration B.

While at this point in time, the use of several movable microreactor cores may appear futuristic, large industrial investments into affordable/smaller scale nuclear technologies offer well-founded hope for the future. 
After 10+ years of the scheduled JUNO run, these new technologies may become a reality to contemplate the measurement scheme proposed in this paper.
The same sources would enable a program beyond $\thone$, including measuring a large sample of $\bar\nu_e$-electron scattering with $\mathcal{O}(10^5)$ events per 100~MW$\cdot$year, searches for eV-sterile neutrino oscillations, and light new particles produced in nuclear and electromagnetic processes. 
We return to these topics in a future study~\cite{HMBP_inprep}.


\begin{acknowledgements}
We thank Stephen Parke for many useful suggestions on this topic. 
We also thank Patrick Huber and Bryce Littlejohn for discussions. 
This work was partially supported by the University of Iowa's Year 2 P3 Strategic Initiatives Program through funding received for the project entitled ``High Impact Hiring Initiative (HIHI): A Program to Strategically Recruit and Retain Talented Faculty.'' 
The work of MP is supported in part by the DOE grant DE-SC0011842.

We acknowledge the use of large language models for coding, research, text editing, and generating components of our figures.
\end{acknowledgements}

\appendix

\section{Oscillations}
\label{app:oscillations}

An exact form for the three-flavor $\nuebar$ survival probability in vacuum is~\cite{Minakata:2007tn}
\begin{multline}
  \label{eq:surv_exact}
  P_{ee} = 1
    - \tfrac{1}{2}\sinsqtwo{\thone}\\
      \Bigl[1 - \sqrt{1-\sinsqtwo{\thtwo}\sin^2\!\Dtwo}\;
        \cos(2|\Dee|\pm\Phi_{\odot})\Bigr] \\
    - \sinsqtwo{\thtwo}\cos^4\!\thone\,\sin^2\!\Dtwo,
\end{multline}
with
\begin{equation}
  \Dee = \frac{\dmee\,L}{4E_\nu},\quad
    \Dtwo = \frac{\dmtwo\,L}{4E_\nu},
\end{equation}
where $\dmee\equiv\cos^2\!\thtwo\,\Delta m^2_{31}+\sin^2\!\thtwo\,\Delta
m^2_{32}$~\cite{Nunokawa:2005nx,Minakata:2006gq,Parke:2016joa} is the effective atmospheric
splitting, and 
\begin{equation}
    \Phi_{\odot}=\arctan(\cos 2\thtwo\,\tan\Dtwo)-\Dtwo\cos 2\thtwo    
\end{equation}
is the phase advancement or retardation of the atmospheric frequency.
The sign in front of $\Phi_\odot$ in \cref{eq:surv_exact} is $+$ve for normal and $-$ve for inverted ordering.

At very short baselines, when the solar phase is small, $\Dtwo\ll1$, we can safely send $\sin^2\!\Dtwo \to 0$ and $\Phi_\odot \to 0$, recovering the two-flavor limit of \cref{eq:surv_exact}. 
At JUNO's standard baseline ($L=52.5\km$, $E\sim3\MeV$) the atmospheric phase $\Dee \simeq 18 \pi$ is large and can be detected as small $\thone$ oscillations with amplitude of $\lesssim2\%$ near the solar-oscillation minimum.
This is a much smaller effect than the $\lesssim 8.7\%$ dip we target in the main text.

In this work, we adopt the NuFit~6.0 normal-ordering central values~\cite{Esteban:2024eli},
\begin{align}
&\sin^2\thtwo=0.307,
\\\nonumber 
&\sin^2\thone=0.0222,
\\\nonumber 
&\dmtwo=7.41\times10^{-5}\eV^2,
\\\nonumber 
&\dmee=2.494\times10^{-3}\eV^2
\end{align}
The value of $\theta_{23}$ is irrelevant for $\nuebar \to \nuebar$ oscillations.

\begin{figure}[t]
\centering
\includegraphics[width=\columnwidth]{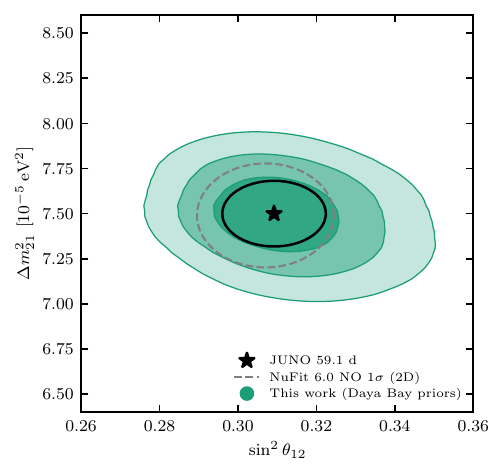}
\caption{Two-dimensional confidence contours ($1,2,3\sigma$, 2 d.o.f.) on $(\sin^2\thtwo,\,\dmtwo)$ for our 59.1-day JUNO simulation using the data release of \cite{JUNO:2025gmd} and the Daya Bay flux model of \cite{DayaBay:2025ngb}. 
Star and ellipse: best fit and $1\sigma$ region from the first JUNO measurement~\cite{JUNO:2025gmd}. 
Dashed grey: NuFit\,6.0 prior~\cite{Esteban:2024eli}. 
The 1D profiled uncertainties agree with the published values at the ${\leq}5\%$ level.
}
\label{fig:validation}
\end{figure}

\section{Reactor Flux Models}
\label{app:flux}

In this appendix, we discuss the details of our reactor flux models used in Configurations A and B.

\subsection{Daya Bay Flux Model}
The Daya Bay data release~\cite{DayaBay:2025ngb} provides the IBD yields per fission, $\Phi^0_i$, in units of $\mathrm{cm^2\,fission^{-1}}$.
It provides $n=25$ neutrino-energy bins in the range $E_\nu \in [1.8, \,9.5]$~MeV with uniform bins of size $0.25$~MeV and one last bin spanning $[7.8,\,9.5]$~MeV. 
This is done for three spectra: \{\uthirtyfive,\pthirtynine, Total\}.
The ``Total" component is defined as the average fuel composition in a cycle, corresponding to fission fractions $f_i(\beta)$ at Daya Bay of $0.564:0.076:0.304:0.056$, with $i=\,^{235}$U, $^{238}$U, $^{239}$Pu, $^{241}$Pu.
Also provided is a $75 \times 75$ covariance matrix $V$ for these three components.
The flux normalization and shape uncertainty in our standalone JUNO analysis uses only the Total $25\times25$ sub-block component to model the $\gwth$-scale far reactors.

\begin{figure}[t]
\centering
\includegraphics[width=\columnwidth]{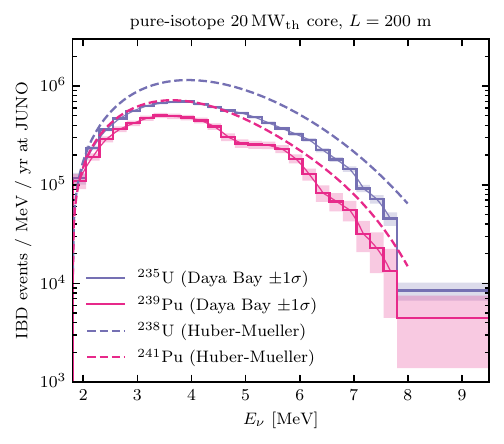}
\caption{
The IBD rate at JUNO for an isotopically pure $20\mwth$ core at 200~m.
We show the measured Daya Bay \uthirtyfive and \pthirtynine yields~\cite{DayaBay:2025ngb} per released bin with their $1\sigma$ uncertainties (the diagonal of the covariance) as bands, together with the linear interpolation between bin centers used in our analysis (thin lines, shown up to the $7.55$--$7.8$~MeV bin center), and the Huber-Mueller~\cite{Huber:2011wv,Mueller:2011nm} \uthirtyeight and \pfortyone spectra (dashed).
The latter are set to zero above the end of their fit range at $8.0$~MeV.
}
\label{fig:isotopic_evolution}
\end{figure}

\Cref{fig:isotopic_evolution} shows the measured \uthirtyfive and $^{239}$Pu yields with their released $1\sigma$ uncertainties (the diagonal of the Daya Bay covariance matrix). 
The yields are converted into IBD rates for JUNO for a hypothetical pure $20\mwth$ core of each isotope at $L=0.2\km$, together with the Huber-Mueller $^{238}$U and $^{241}$Pu spectra used for the minority components~\cite{Huber:2011wv,Mueller:2011nm}, valid only up to $8.0$~MeV and taken as zero above this point for simplicity.

\subsection{Simplified Fuel Depletion Model}
\label{app:depletion}

For the Configuration A microreactor studies, we assume a small HALEU reactor with high \uthirtyfive isotopic composition. 
Given a HALEU core of thermal power $P$ with fission fractions $f_i(t)$, we evolve its composition with a simple fuel depletion model.
In general, this problem can be quite complicated, but we make the following key assumptions: i) the neutron spectrum is thermal and fixed throughout the entire lifetime of the reactor core, ii) the core starts fresh with a dominant \uthirtyfive and subdominant \uthirtyeight populations, iii) the thermal power is exclusively set by the fission energy release and is held constant over the operation of the reactor.
Note that for our setup, only the drift of the fuel composition between stops matters.
Any time-independent feature of the spectrum is common to every stop and is calibrated by the near-site reactor data and flux monitors.

\paragraph{Evolution equations.}
Under these assumptions the evolution of the number of atoms $N_i$ in the core is
\begin{align}\label{eq:bateman}
    \dot N_{235} &= - \phi\, \Sigma_{235}\, N_{235},
    \nonumber\\
    \dot N_{238} &= - \phi\, \Sigma_{238}\, N_{238},
    \nonumber\\
    \dot N_{239} &= \phi\, \langle \sigma^{238}_{n,\gamma}\rangle N_{238} - \phi\, \Sigma_{239}\, N_{239},
    \\
    \dot N_{240} &= \phi\, \langle \sigma^{239}_{n,\gamma}\rangle N_{239} - \phi\, \Sigma_{240}\, N_{240},
    \nonumber\\
    \dot N_{241} &= \phi\, \langle \sigma^{240}_{n,\gamma}\rangle N_{240} - \left(\phi\, \Sigma_{241} + \lambda_{241}\right) N_{241},
    \nonumber
\end{align}
where $\phi$ is the total neutron flux, and $\langle\sigma^i_{n,f}\rangle$ and $\langle\sigma^i_{n,\gamma}\rangle$ are the fission and capture cross sections averaged over the (fixed) neutron spectrum.
The total depletion cross sections are $\Sigma_i=\langle\sigma^i_{n,\gamma}\rangle+\langle\sigma^i_{n,f}\rangle$.
For \uthirtyeight we account only for capture, $\Sigma_{238}=\langle\sigma^{238}_{n,\gamma}\rangle$.
\uthirtyeight fissions only with fast neutrons.
Since the fast flux is set by the total fission rate, which is fixed at constant power, and $N_{238}$ barely changes, its fission fraction remains roughly constant at its initial value.
Like \uthirtyeight, ${}^{240}$Pu is fertile rather than fissile, and we neglect its thermal fission, so $\Sigma_{240} = \langle\sigma^{240}_{n,\gamma}\rangle$.
The $\beta$ decay of \pfortyone to ${}^{241}$Am, with $\lambda_{241}=\ln 2/14.3\,\text{yr}$, is not negligible over a decade and is kept.

The fresh core is fixed by the uranium mass $M_{\rm U}$ and enrichment $w$,
\begin{align}
    N_{235}(0) &= \frac{w\, M_{\rm U}}{235\text{ g/mol}}N_A,
    \\\nonumber
    N_{238}(0) &= \frac{(1-w)\, M_{\rm U}}{238\text{ g/mol}}N_A,
    \\\nonumber
    N_{239}(0) &= N_{240}(0) = N_{241}(0) = 0.
\end{align}
Since the neutron spectrum is held fixed, the coefficients in \cref{eq:bateman} are constant when written in terms of the neutron fluence $u(t)=\int_0^t \phi(t')\,dt'$.
In this case, the isotope numbers are given by the closed form
\begin{align}
    N_{i}(u) &= N_{i}(0)\, e^{-u\Sigma_i}, \qquad i\in\{235,\,238\},
    \nonumber\\
    N_{239}(u) &= \frac{\langle \sigma^{238}_{n,\gamma}\rangle N_{238}(0)}{\Sigma_{239} - \Sigma_{238}} \left(e^{-u\Sigma_{238}} -e^{-u\Sigma_{239}}\right).
\end{align}
The higher isotopes ${}^{240}$Pu and \pfortyone follow from the same chain.
Because of the decay term, \pfortyone does not have a simple closed form in $u$, and we solve the full chain numerically.
The relation between $u$ and $t$ then follows from the condition for constant power $P=\phi\sum_i E_i\langle\sigma^i_{n,f}\rangle N_i$.
If power is to remain constant in our model, the neutron flux $\phi$ should slowly rise as \uthirtyfive is burnt inside the core.
The thermal fission fractions are then $f_i \propto \langle\sigma^i_{n,f}\rangle N_i$ for $i\in\{235,239,241\}$.

We use representative values for the cross sections from ENDF/B-VIII.0~\cite{Brown:2018jhj} for a thermal spectrum representative of a graphite-moderated core, as listed in \cref{tab:onegroup}.

\begin{table}[t]
\centering
\begin{tabular}{lccccc}
\toprule
 & \uthirtyfive & \uthirtyeight & \pthirtynine & ${}^{240}$Pu & \pfortyone \\
\midrule
$\langle\sigma_{n,f}\rangle$      & 136~b & --   & 410~b & --    & 393~b \\
$\langle\sigma_{n,\gamma}\rangle$ & 29~b  & 2.6~b & 240~b & 460~b & 147~b \\
\bottomrule
\end{tabular}
\caption{Representative values for the one-group cross sections for the HALEU core.}
\label{tab:onegroup}
\end{table}

\paragraph{Benchmark core.}
We take a $20\mwth$ core loaded with $M_{\rm U}=1.52$~t at $w=19.75\%$.
With our evolution parameters, about $3\%$ of the \uthirtyfive inventory is consumed per year.
The key quantity that determines amount of plutonium bred in the core (and therefore the spectral changes over time) is the initial conversion ratio
\begin{equation}
    \mathrm{CR}_0=\frac{\langle\sigma^{238}_{n,\gamma}\rangle N_{238}(0)}{\Sigma_{235}N_{235}(0)}=0.063.
\end{equation}
The typical value above is much smaller than in a commercial core because the ratio $N_{238}/N_{235} \simeq 4$ for a HALEU core with $19.75\%$ enrichment is much smaller than the typical $N_{238}/N_{235} \simeq 24$ for a LEU core at $4\%$ enrichment.
Note that in principle $\mathrm{CR}_0$ depends on the neutron spectrum through the effective \uthirtyeight capture cross section.

With the above inputs, \pthirtynine goes from a negligible isotope to reaching a fission fraction of a few percent after about ten years.
Its growth rate is initially $\simeq0.5\%$ per year and slows as \pthirtynine approaches its equilibrium abundance.
Since $f_{239} \propto \mathrm{CR}_0$ to a good approximation, we test the dependence on the spectrum by varying $\mathrm{CR}_0$ (i.e., $\langle\sigma^{238}_{n,\gamma}\rangle$) by a factor of two in either direction.
We find that our results are no more sensitive to this choice than to the reproducibility uncertainty $\sigma_P$.

\paragraph{Fission fractions used in the analysis.}
We take the fission fractions of the HALEU core directly from the depletion chain of \cref{eq:bateman}.
We solve it numerically at constant power, with the flux fixed at each time step by $P=\phi\sum_i E_i\langle\sigma^i_{n,f}\rangle N_i$.
The thermal fission fractions $f_i \propto \langle\sigma^i_{n,f}\rangle N_i$ are normalized to $1-f_{238}$, and the fast-fission fraction of \uthirtyeight is held fixed at $f_{238}=0.02$.

For a fresh core, the starting point is $f_{235}=98\%$ and $f_{238}=2\%$.
After five years the chain gives $f_{239}=2.5\%$ and $f_{241}=0.04\%$.
After the ten years of Configuration~A it gives $f_{239}=4.7\%$, $f_{241}=0.3$\% and $f_{235}=93\%$.
\pfortyone remains a per-mille effect throughout, since it requires two further captures beyond \pthirtynine and decays on a timescale comparable to the program.
The fuel composition of the movable source is averaged over each stop in $60$~d steps.

\begin{figure*}[t]
\centering
\includegraphics[width=\columnwidth]{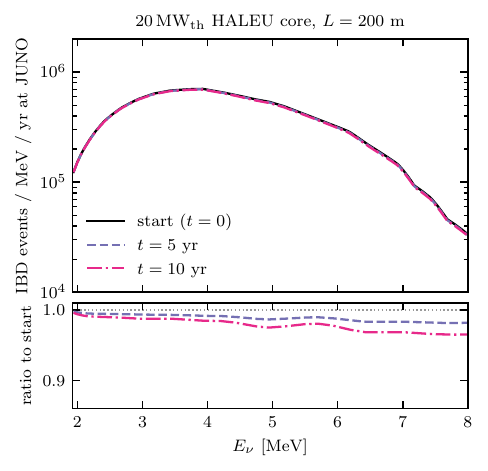}
\includegraphics[width=\columnwidth]{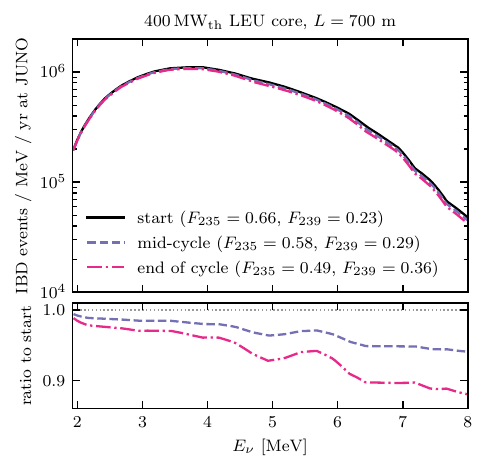}
\caption{
Left: the $20\mwth$ HALEU core at 200~m at the start, after 5~years and after 10~years of operation (\cref{eq:bateman}), built from the spectra of \cref{fig:isotopic_evolution}.
Right: the $400\mwth$ LEU core at 700~m at the Daya Bay measured fission fractions~\cite{DayaBay:2017jkb} at the start, middle and end of its 18-month cycle.
Both show unoscillated IBD events at JUNO at fixed thermal power; the bottom panels show the ratio to the start spectrum.
\label{fig:full_evolution}
}
\end{figure*}

\paragraph{Resulting spectra.}
The left panel of \cref{fig:full_evolution} shows the HALEU core at the start, after $5$~years, and after $10$~years, at fixed thermal power.
After $10$~years the IBD rate per unit power is $1.8\%$ lower, and the flux near $8$~MeV is $3.5\%$ lower.
\uthirtyfive still produces about $93\%$ of the detected $\nuebar$ at the end of the program.
For comparison, the right panel shows a commercial LEU core at $400\mwth$ and $L = 700$~m, following the fission fractions measured by Daya Bay over an 18-month cycle~\cite{DayaBay:2017jkb} ($F_{235}=0.66\to0.49$, $F_{239}=0.23 \to 0.36$).
These are representative of an equilibrium PWR cycle, and we use them directly for the LEU core of Configuration~B.
At equal thermal power and baseline, the LEU core yields $5\%$ fewer IBD events than the HALEU core already at the start of its cycle, because its fuel contains a sizable plutonium fraction, and plutonium releases fewer $\nuebar$ per unit of thermal energy than \uthirtyfive.
Over a single 18-month cycle, the LEU rate drops by a further $5\%$ and its flux near $8$~MeV by $12\%$.
The HALEU core takes ten years to change by much less, dropping only $1.8\%$ in the total IBD rate and $3.5\%$ near $8$~MeV region.

\section{Validation with JUNO's First Results}
\label{app:validation}

We now describe our study to reproduce JUNO's oscillation results, following the methods outlined in the JUNO article and data release~\cite{JUNO:2025gmd} as well as in NuFit's analysis~\cite{Esteban:2026phq}.
A key systematic uncertainty in this analysis is that on the reactor flux predictions. 
We follow JUNO and use Daya Bay's flux unfolding procedure as described in \cite{DayaBay:2021dqj} and \cite{Esteban:2026phq}.

Following Appendix~A of the NuFit analysis~\cite{Esteban:2026phq}, the continuous flux times cross section is written as,
\begin{equation}
\label{eq:appA_shape}
S(E)=\phi_{\rm HM}(E)\,\sigma_{\rm IBD}(E)\;\sum_{m=1}^{n}y_m\,\delta_m(E),
\end{equation}
where $\phi_{\rm HM}$ is the Huber-Mueller spectrum at the Daya Bay average fission fractions (with no isotope correction) and $\{\delta_m\}$ are the cardinal cubic splines through the 25 bin centers ($\delta_m(E_{m'})=\delta_{mm'}$).  
The coefficients are fixed by the measured yields at JUNO, that is,
\begin{equation}
    M_{im}=\frac{1}{\Delta E_i}\int_{\rm bin\,i}\phi_{\rm HM}\sigma_{\rm IBD}\,\delta_m\,dE,\qquad
\bm y^0=M^{-1}\bm\Phi^0.
\label{eq:appA_M}
\end{equation}
While ad-hoc, this procedure accounts for the exact fuel composition model of the far reactors that we do not have access to.

The covariance is recast as $n$ Gaussian pulls by eigendecomposition, $V=O\,D\,O^T$, $\Psi=O\sqrt D$ (columns are modes), and each mode is pushed through the same interpolation, $\bm y^{(k)}=M^{-1}\Psi_{\cdot k}$, so that a pull $\xi_k$ distorts the flux by
\begin{equation}
S(E) \to S(E)\left[1 + \sum_k \xi_k \, \frac{\sum_n y^{(k)}_n \delta_n(E)}{\sum_n y^0_n \delta_n(E)} \right], \quad \chi^2\supset\sum_k\xi_k^2.
\label{eq:appA_pulls}
\end{equation}
Note that the pull penalty is diagonal.  
For the near-reactor, the same construction is applied to the relative covariance $V_{ij}/(\Phi^0_i\Phi^0_j)$ of the \uthirtyfive block.
The modes are then interpolated linearly in $E$ and enter the Asimov covariance $C=\mathrm{diag}(\mu)+\sum_kv_kv_k^T$ as $v_k=t\,\partial\mu/\partial\xi_k$.
Modes with eigenvalue below $10^{-10}$ of the largest are dropped.

For the standard far-reactor analysis, the NuFit prescription additionally rescales the
prediction bin-per-bin to JUNO's recovered un-oscillated spectrum,
\begin{equation}
k_i=\frac{\big(N^{\rm obs}_i-B_i\big)/\overline{P}_{ee,i}}{N^{\rm pred,unosc}_i},
\label{eq:binscale}
\end{equation}
where $\overline{P}_{ee,i}$ is JUNO's own released model survival probability per bin;
this pins the un-oscillated shape to data and turns flux-model and cross-section choices
into nulls (it was found to be the decisive ingredient in reproducing JUNO's solar-parameter
result to $0.06\sigma$).  It has no analogue for projections, where the measured flux with
its covariance takes its place.

The test statistic is defined as
\begin{align}
\chi^2 &    = \sum_i\frac{\big(N^{\rm obs}_i-N^{\rm pred}_i(\bm\xi)\big)^2}{\sigma^2_{{\rm CNP},i}}
+\sum_a\Big(\frac{\xi_a}{\sigma_a}\Big)^2,
\label{eq:chi2}
\end{align}
with $(\sigma^2_{{\rm CNP},i}/3)^{-1} = (N^{\rm obs}_i)^{-1}+(N^{\rm pred}_i/2)^{-1}$ following the combined Neyman-Pearson~\cite{Ji:2019yca} form used by JUNO.
The predicted number of events in a given bin is given by
\begin{align}
\label{eq:pred}
  N^{\rm pred}_i &= (1+\xi_{\rm norm})\,k_i\,\mathcal N\sum_{\rm cores\,c}\Bigg[R\big(\xi_{\rm scl},\xi_{\rm bias},\xi_{\rm res}\big) \times
\\\nonumber
  & S\big(E;\{\xi_k\}\big) \,\frac{P_{ee}(E,L_c)}{4\pi L_c^2}\Bigg]_i 
  \\\nonumber
  &+\sum_bB_{b,i}\,(1+\xi_b)\,(1+\delta_{b,{\rm LiHe}}\,\xi_{\rm sh}E_i),
\end{align}
with the 9 reactor cores (Yangjiang, Taishan, and the effective Daya Bay core at 215~km).
Matter effects are included with a matter density of $2.6\,\mathrm{g\,cm^{-3}}$~\cite{Li:2016txk,Khan:2019doq}.

With this treatment our analysis reproduces NuFit's Tab.~2 relative fluxes and best fit well.
Our analysis also reproduces JUNO's official $(\sin^2\theta_{12},\dmtwo)$ to less than $0.1\sigma$.
We find that removing Fangchenggang reactors at 412~km from the ``world reactors" backgrounds and including it as an oscillated signal does not meaningfully improve our agreement with JUNO's official results.  
The most critical factors in reproducing JUNO's numbers were the bin-to-bin rescaling of \cref{eq:binscale}, the $2.4\%$ normalization, and including the Daya Bay complex in the signal model.
Note that the Daya Bay reactor antineutrinos undergo almost a full oscillation cycle on the way to JUNO, $\Delta_{21}^{\rm Daya Bay} \simeq 2\pi$, arriving close to a $\nuebar$ survival maximum.
This happens in the same energy range where the $53$~km Taishan and Yangjiang flux sits at its first solar minimum, $\Delta_{21}\simeq\pi/2$, so Daya Bay partially fills the oscillation dip and should not be neglected.

\bibliographystyle{utphys}
\bibliography{main}

\end{document}